\documentstyle[12pt,aaspp4]{article}
\def\plotone#1{\centerline{\epsfxsize=5.0in\epsfbox{#1}}}
\def\BE{\begin{equation}}
\def\BEL#1{\begin{equation}\label{#1}}
\def\EE{\end{equation}}
\newcommand{\etal}{et al.}
\newcommand{\HI}{H\,{\scriptsize I}}
\newcommand{\HII}{H\,{\scriptsize II}}
\newcommand{\LOGTEN}{{\log_{10}}}
\newcommand{\vLSR}{{v_{\rm LSR}}}
\newcommand{\Hplus}{{\rm H}^+}
\newcommand{\Bband}{B}
\newcommand{\Vband}{V}
\newcommand{\Rband}{R}

\newcommand{\Rkc}{R_{\rm KC}}
\newcommand{\Ikc}{I_{\rm KC}}
\newcommand{\Rspin}{R_{\rm S}}
\newcommand{\UBVRIkc}{UBV\Rkc\Ikc}
\newcommand{\griz}{griz}
\newcommand{\ugriz}{u'g'r'i'z'}
\newcommand{\ubvby}{ubv{\beta}y}
\newcommand{\JHKLp}{JHKL'}
\newcommand{\BminusV}{({\Bband}{\rm -}{\Vband})}
\newcommand{\BminusR}{({\Bband}{\rm -}{\Rband})}

\newcommand{\Ebv}{E\BminusV}
\newcommand{\EbvBH}{E^{\rm BH}} 
\newcommand{\Ebr}{E\BminusR}
\newcommand{\Eby}{E(b{\rm -}y)}
\newcommand{\Rv}{R_V}
\newcommand{\Ab}{A(\Bband)}
\newcommand{\Av}{A(\Vband)}

\newcommand{\Mgtwo}{{\rm Mg}_2}
\newcommand{\fcut}{f_{\rm cut}}
\newcommand{\lameff}{\lambda_{\rm eff}}
\newcommand{\AoverAv}{A/A(\Vband)}
\newcommand{\sigBV}{\sigma_{\rm BV}} 
\newcommand{\sigDI}{\sigma_{\rm D-I}} 
\newcommand{\RDmap}{{\bf D}} 
\newcommand{\LDmap}{{\bf D}^L} 
\newcommand{\QDmap}{{\bf D}^Q} 
\newcommand{\FDmap}{{\bf D}^S} 
\newcommand{\Kmap}{{\bf K}} 
\newcommand{\Xmap}{{\bf X}} 
\newcommand{\Tmap}{{\bf T}} 
\newcommand{\Rmap}{{\bf R}} 
\newcommand{\NDmap}{{\bf D}^T} 
\newcommand{\Hmap}{{\bf H}} 
\newcommand{\Wmap}{{\bf W}} 
\newcommand{\sigmap}{{\bf \sigma}} 
\newcommand{\Ides}{{\bf I}_{des}} 
\newcommand{\Icorr}{{\bf I}_{corr}} 
\newcommand{\Wsq}{W^\Box(21\arcmin)} 
\newcommand{\Wg}{W^G} 
\newcommand{\Diffmap}{{\bf S}} 
\newcommand{\Ang}{{\rm ~\AA}}

\newcommand{\degree}{^\circ}
\newcommand{\kms}{{\rm ~km/s}}
\newcommand{\Kkms}{{\rm ~K~km/s}}
\newcommand{\MAG}{{\rm ~mag}}
\newcommand{\cm}{{\rm ~cm}}
\newcommand{\Jy}{{\rm ~Jy}}

\newcommand{\MJypSr}{{\rm ~MJy}/{\rm sr}}
\newcommand{\GHz}{{\rm ~GHz}}
\newcommand{\K}{{\rm ~K}}
\newcommand{\pc}{{\rm ~pc}}
\newcommand{\nWpMMSr}{{\rm ~nW/m}^2/{\rm sr}}
\newcommand{\sqasec}{\hbox{\rlap{$\sqcap$}$\sqcup$}}
\newcommand{\magpasec}{{\rm ~mag} / \sqasec}
\newbox\grsign \setbox\grsign=\hbox{$>$} \newdimen\grdimen \grdimen=\ht\grsign
\newbox\simlessbox \newbox\simgreatbox
\setbox\simgreatbox=\hbox{\raise.5ex\hbox{$>$}\llap
     {\lower.5ex\hbox{$\sim$}}}\ht1=\grdimen\dp1=0pt
\setbox\simlessbox=\hbox{\raise.5ex\hbox{$<$}\llap
     {\lower.5ex\hbox{$\sim$}}}\ht2=\grdimen\dp2=0pt

\def\simlt{\mathrel{\copy\simlessbox}}

\begin{document}

\title{Maps of Dust IR Emission for Use in Estimation of Reddening 
and CMBR Foregrounds}

\author{David J. Schlegel}
\affil{University of Durham, Department of Physics,
 South Road, Durham DH1 3LE, United Kingdom}
\authoremail{D.J.Schlegel@durham.ac.uk}
\and
\author{Douglas P. Finkbeiner \& Marc Davis}
\affil{University of California at Berkeley, Departments of Physics and 
Astronomy, 601 Campbell Hall, Berkeley, CA 94720}
\authoremail{dfink@astro.berkeley.edu, marc@coma.berkeley.edu}


\begin{abstract}

We present a full sky $100\micron$ map that is a reprocessed composite
of the COBE/DIRBE and IRAS/ISSA maps, with the zodiacal foreground and  
confirmed point sources removed.
Before using the ISSA maps, we remove the remaining artifacts from
the IRAS scan pattern.  Using the DIRBE $100\micron$ and $240\micron$ data, 
we have constructed a map of the dust temperature, 
so that the $100\micron$ map
may be converted to a map proportional to dust column density.  The dust
temperature varies from $17\K$ to $21\K$, which is modest but
does modify the estimate of the dust column by a factor of 5.
The result of these manipulations 
is a map with DIRBE-quality calibration and IRAS
resolution.  A wealth of filamentary detail is
apparent on many different scales at all Galactic latitudes.
In high latitude regions, 
the dust map correlates well with maps of \HI\ emission, but deviations
are coherent in the sky and are especially conspicuous in regions of saturation 
of \HI\ emission toward denser clouds and the formation
of ${\rm H}_2$ in molecular clouds.  In contrast, high-velocity \HI\ clouds are
deficient in dust emission, as expected.

To generate the full sky dust maps, we must first remove zodiacal light
contamination as well as a possible cosmic infrared background 
(CIB).  This is done via a regression analysis of the $100\micron$ DIRBE
map against the Leiden-Dwingeloo map of \HI\ emission, with corrections for
the zodiacal light via a suitable expansion of the DIRBE $25\micron$ flux.
This procedure removes virtually all traces of the zodiacal
foreground.  For the $100\micron$ map no significant CIB is detected.
At longer wavelengths, where the zodiacal
contamination is weaker, we detect the CIB at surprisingly high
flux levels of $32 \pm 13 \nWpMMSr$ at $140\micron$, and $17 \pm 4 \nWpMMSr$
at $240\micron$ ($95\%$ confidence).  This integrated flux is $\sim 2$
times that extrapolated from optical galaxies in the Hubble Deep Field.

The primary use of these maps is likely to be as a new estimator of
Galactic extinction.  To calibrate our maps, we assume a standard reddening
law, and use the colors of elliptical galaxies to measure the reddening
per unit flux density of $100\micron$ emission.  We find consistent calibration
using the $\BminusR$ color distribution of a sample of 106 brightest
cluster ellipticals, as well as a sample of 384 ellipticals with $\BminusV$
and Mg line-strength measurements.  For the latter sample, we use the
correlation of intrinsic $\BminusV$ versus $\Mgtwo$ index to greatly
tighten the power of the test.  We demonstrate that the new maps are twice
as accurate as the older Burstein-Heiles reddening estimates in regions of
low and moderate reddening.  The maps are expected to be significantly more
accurate in regions of high reddening.  These dust maps will also be useful
for estimating millimeter emission that contaminates CMBR experiments and
for estimating soft X-ray absorption.

We describe how to readily access our maps for general use.

\end{abstract}


\keywords{Cosmology: observations --- dust --- extinction ---
 infrared: ISM: continuum}


\section{Introduction}

In the past 15 years, two NASA missions have revolutionized our
knowledge of the diffuse interstellar medium.  The path-breaking
Infrared Astronomy Satellite (IRAS) mission of 1983 led to the first
full-sky maps of the diffuse background radiation in four broadband
infrared channels, centered at 12, 25, 60, and $100\micron$, with a $\sim 5$
arcminute beam.  The DIRBE experiment (Diffuse InfraRed Background
Experiment) onboard the COBE satellite imaged the full sky in 10 broad
photometric bands from $1\micron$ to $240\micron$ with a beam of
$0.7\degree$.  This experiment, for all but the shortest wavelength
channels, was active for 42 weeks in 1989-1990 before its $^4{\rm
He}$ cryogen was exhausted.  Although IRAS was optimized to detect
point sources and sources of small angular extent (Beichman \etal\ 
1988), it has been possible to create large area sky maps from the
IRAS data stream (ISSA images: \cite{issa94}).  Striping artifacts
from time variation in the zodiacal foreground emission has largely
been filtered out of the individual ISSA maps, but artifacts remain
and the zero-point has large-scale drifts across the images.
The DIRBE experiment had a much
better control of the absolute calibration and has two channels further to
the submillimeter than does IRAS (at $140\micron$ and $240\micron$),
but the maps have much less angular resolution.

The diffuse emission seen in these experiments is a superposition of
zodiacal foreground emission, dust and molecular emission from the
interstellar medium (ISM), point sources from within and beyond the
Milky Way, and possibly a diffuse extragalactic background.
Separating these components has proven to be a most difficult task
(\cite{hauser96}). The zodiacal emission dominates the $12\micron$
and $25\micron$ channels and is a quite serious contaminant at
$60\micron$, while its contribution to the $100\micron$ maps is
less strong.  Modelling of the zodiacal
emission has proven quite complex (\cite{reach96}).  One must consider
that the zodiacal dust in the ecliptic plane has a distribution of
temperatures, that the tilt of Earth's orbit relative to the midplane
of the dust is readily detectable, and that the zodiacal emission
shows a strong dependence on solar elongation, information which is
lost if one uses only the annual-averaged map of the DIRBE data.

A full-sky map of the diffuse radiation serves many purposes, and we
here report our efforts to combine the $100\micron$ maps of IRAS and
DIRBE in a manner designed to be accessible to the general
astrophysical community.  To construct this map, we have removed the
zodiacal foreground emission from the DIRBE maps, removed striping
artifacts from the IRAS/ISSA maps, subtracted confirmed point sources,
and combined the maps in such a way as to preserve the DIRBE calibration
and IRAS resolution.  We have used the ratio of $100\micron$ to
$240\micron$ emission to deduce a dust color temperature, allowing us to
translate the $100\micron$ emission to a column density of radiating
dust.  Such a procedure may be inadequate towards complex, dense
clouds at low Galactic latitude, but in most directions the emission
is dominated by dust within a single environment and radiation field.

Since the diffuse emission in the infrared is a direct measure of the
column density of the interstellar dust, such a map can be used as a
measure of extinction for extragalactic objects.  As we shall argue
below, the new dust map has better angular resolution and better
control of systematics than possible with the reddening maps of
Burstein \& Heiles (1978, 1982).
The essence of the Burstein-Heiles procedure was to assume that variations
in the dust-to-neutral-gas ratio can be adequately modelled by the smoothed 
mean background of galaxies (at northern declinations), or that this
ratio is constant (at southern declinations).
We make no assumptions about variations
in dust-to-gas; neither do we need make any assumption about the
correlation of diffuse ionized gas with neutral hydrogen.
We only make the weaker assumption
that the distribution of dust grain sizes is everywhere the same,
since the relationship between UV/optical extinction and far-IR emission
depends on the grain size distribution
(\cite{draine84}; \cite{draine85}; \cite{guhathakurta89}; \cite{mathis92}).
The diffuse gas, where our results apply, is thought to have reasonably uniform
dust properties, with values of $\Rv \equiv \Av/\Ebv \approx 3.1$.
Since neither the
Burstein-Heiles nor our assumptions are true for all environments, a
comparison between the different methods is of interest.

To calibrate the extinction curve, we extensively explored the
anticorrelation of counts-in-cells of the APM galaxy survey (Maddox \etal\ 
1990b, 1990c) with $100\micron$ emission, finding a consistent
calibration in different directions of the sky.  Due to complications
in translating this calibration into $\Ebv$, we shall defer this
discussion for a separate paper (\cite{finkbeiner97}).

Since we find that the $100\micron$ maps are well correlated
with \HI\ emission at high latitudes,
we do not not expect the new extinction maps
to grossly deviate from the Burstein-Heiles reddening estimates
when averaged to the same scales.
But many dusty regions have filamentary structure with large
fluctuations in extinction estimates over angular scales much smaller
than the resolution of the BH maps.  Furthermore, at low
Galactic latitudes, regions such as Orion and Ophiuchus might be saturated
in \HI\ emission, or partially ionized, whereas the dust remains optically thin.

Another important application for maps of the diffuse emission is to
compare with CMBR fluctuation maps.  Recent analyses of the COBE/DMR
fluctuations (\cite{kogut96}) show a positive correlation with DIRBE
maps, particularly in the North Polar Spur region.  Similar
correlations are seen between the Saskatoon CMBR maps and the IRAS maps
(\cite{oliveira97}; \cite{leitch97}; \cite{jaffe97}).
This is unexpected, since these $40 \GHz$ CMBR experiments are not sensitive to
dust emission, but are sensitive to free-free emission from the Warm Ionized
Medium (WIM).  This implies that the WIM is at least partly correlated
with the cold, dusty medium, and that our dust maps could be used as
a model for both components (\cite{reynolds95}).
Alternatively, the observed CMBR correlations might be explained by
small spinning dust grains producing continuum emission in the 
$10-100\GHz$ range (\cite{draine97}), in which case the dust maps should
correlate extremely well with CMBR observations over the entire sky.

This paper is organized as follows: Section \ref{section_column}
describes the processing of the DIRBE data.  We describe how the
weekly averaged DIRBE maps are trimmed and then reconstituted into
annual average maps more suitable for our simple treatment of the
zodiacal light removal.  We fit a zodiacal light model by maximizing
the correlation between the DIRBE maps and \HI\ in regions of high
Galactic latitude and low flux.

The diffuse cirrus emission is expected to have a reasonably uniform
temperature since the ultraviolet radiation field which is heating the
dust must be a smooth function of angular position except within dense
clouds.  Fortunately, the emission at $100$ and $240\micron$ is
expected to be dominated by grains sufficiently large to be in
equilibrium with the radiation field (e.g.\ \cite{draine85};
\cite{guhathakurta89}), so that the $100\micron/240\micron$ color ratio can
provide a useful measure of grain temperature.  In Section
\ref{section_tempmap}, we describe how we measure the temperature in
the presence of substantial noise in the $240\micron$ map.  This
information is used to translate the $100\micron$ emission to dust
column density.

The inclusion of the IRAS data is discussed in Section \ref{section_iras}.
Before we can combine the IRAS/ISSA map with the DIRBE map,
we must first deal with the ISSA boundaries and discontinuities.  The
ISSA maps are destriped versions of the original IRAS scans, but further
destriping is possible and essential for high angular-resolution studies.
We use a matched set of window functions that low-pass filter the DIRBE
map to $1\degree$ resolution and high-pass filter IRAS/ISSA maps 
on the same scale.  The combined map is then simply the sum of these two
filtered maps.  Section \ref{section_sources} details the removal of
extragalactic objects and point sources from the maps, leaving behind
only a map of the dust emission.

In Section \ref{section_calibrate}, we present a simple procedure for
normalizing the reddening per unit flux density of dust emission.  We
correlate the residual of the $\BminusV$ vs.\ $\Mgtwo$ index for 384
elliptical galaxies against the estimated reddening, and adjust the
normalization until the slope of the relation is zero.  The resulting
calibration has an uncertainty of only $10\%$.  We also show that the
maps have an accuracy of $16\%$ in predicting reddening, which is twice
as good as the older BH procedure.

A summary of all these reprocessing procedures in presented
in Section \ref{section_psummary}.
Readers uninterested in the details of the analysis 
should skip directly to this point.  General
discussion and conclusions are presented in Sections \ref{section_discussion}
and \ref{section_conclude}.
Appendix \ref{section_contam} discusses the contamination
level of stars that remain in the maps.
Appendix \ref{section_filters} describes how to relate the reddenings
and extinction in different optical and infrared filters.
Appendix \ref{section_files} describes how to access the final maps.


\section{Dust Column Densities From the DIRBE Maps}
\label{section_column}

\subsection{The DIRBE Data Set}

The DIRBE data set is in the public domain and was obtained from the
National Space Science Data Center (\cite{nssdc}) on CDROM. 
For a complete
description of the DIRBE instrument, see Boggess \etal\ (1992). Both
Annual Average Skymaps and maps averaged over each week of the mission
are provided.  The data are binned to $393{,}216$ $0.\degree 32 \times
0.\degree 32$ pixels on the COBE Quadrilateralized Spherical Cube.
(For a description of this ``skycube'' projection, see \cite{chan75}.)
As described below, we create new annual average $25\micron$ and $100\micron$
skymaps  from the
41 weekly maps in skycube format in which the weighting function of
each week is forced to be the same for both channels.  We then 
reproject the data to
standard, equal-area, polar projections oversampled to 512 pixels from
$b=0\degree$ to $b=90\degree$ ($0.0250$ square degrees/pixel).
In order to preserve total flux to a good approximation, we initially
projected to a Lambert (polar) mapping with $4\times4$ smaller pixels which
were then binned to the $0.0250$ square-degree resolution.
(Note that the surface brightness of the cirrus is preserved in this
re-projection.)
These maps for each passband, $b$, are denoted by $\RDmap_b$, and are in
units of $\MJypSr$.  We use the $25\micron$ map as a model of the zodiacal
light, and the $100\micron$ and $240\micron$ maps for measuring the
temperature and column density of the dust.

\subsection{Zodiacal Light Removal}

The DIRBE maps are severely contaminated by zodiacal
emission from dust in the solar system, otherwise
known as the IPD (inter-planetary dust).
The IPD primarily re-radiates thermally at $\sim 280\K$ in the infrared
(for $\nu^0$ emissivity).
Its emission per unit column density is substantially higher than that
of $20\K$ Galactic dust:
using an $\alpha=2$ emissivity law, the $100\micron$ emission of the IPD
is larger by a factor $\sim 10^5$
than that of Galactic dust for an equivalent column density.
The brightest zodiacal emission is in the ecliptic plane, at
the level of $\sim 10 \MJypSr$.  Based upon our solutions,
this corresponds to an extinction
at the negligible level of $\Ab \approx 10^{-6} \MAG$.
Thus, the removal of zodiacal light emission is critical
as it makes no measurable contribution to optical extinction.

The $25\micron$ maps, where the IPD emission peaks, is used to model
the IPD contamination at the longer wavelengths.  Before removing this
contamination, the DIRBE annual average data must first be regenerated
so that the data at all wavelengths are sampled at the same time
and through the same IPD.
The zero-points of the
zodiacal contamination are constrained by \HI\ maps at high Galactic
latitude.

\subsubsection{Dependence on Time and Solar Elongation Angle}

Removal of the zodiacal light is a difficult problem.
The observing strategy of the COBE satellite resulted in a lower
signal-to-noise ratio in the ecliptic plane where the zodiacal light is most
prominent.  The number of high-quality DIRBE observations per pixel
ranged from $\sim 200$ at the ecliptic equator to $\sim 800$ near the
ecliptic poles.  The least-sampled regions were approximately cones of
$90 \degree$ opening angle centered at ecliptic $(\lambda,\beta) = (100\degree,
0\degree)$, and $(\lambda,\beta) = (280\degree, 0\degree)$.  The DIRBE maps
exhibit a striping pattern perpendicular to the ecliptic plane; scans
adjacent in the sky correspond to temporally separated instrument
tracks, taken at varying solar elongation angles, through different
column densities of IPD.  

Measurements for each week of the mission
were used to construct weekly skymaps.  The DIRBE Annual Average
Skymaps are a combination of the weekly maps, averaged by the number
of times that each pixel was observed in a weekly map.
Unfortunately, the whole sky is not scanned each week, and some
parts of the scanned region are omitted in certain weeks due to
interference from the moon and planets.  Furthermore, not all the
detectors were on at all times.

A cursory glance at the weekly skymaps reveals a serious problem with
combining data in this way.  Near the ecliptic plane, the $25\micron$ flux
exhibits a strong dependence on solar elongation angle, $e$.  During the
41 weeks, data were typically taken between $e=64\degree$ and $e=124\degree$,
and the $25\micron$ flux varies by a factor of two over this range.
Comparison with the $100\micron$ IPD flux also reveals a strong temperature
gradient with respect to $e$.  This is easily explained by assuming
a radial gradient of IPD density from the sun, because a
line of sight passing close to the sun must traverse more (and hotter)
dust than a line of sight at larger solar elongation.
As a further complication, because the 10 passbands were not always observed
simultaneously,  the weekly skymaps of the weighting function do not always
match across all channels.  Contributions to the
Annual Average Skymaps were weighted according
to the number of observations made in each channel, 
in spite of the strong dependence on elongation angle.  

In cases where the measurement noise is overwhelmed
by the time-dependent systematic variations, this method of averaging
is inappropriate.
The weights assigned to sky pixels are thus not always the same for
the $25\micron$ and $100\micron$ weekly maps, leading to weighted Annual
Average Skymaps which cannot be directly compared to each other.  
To rectify this, we recombine the 41 weekly skymaps using the same weight
for the same pixels in different passbands.  
We use the error maps for the $100\micron$ passband
as weights in all other passbands.  
Because the IPD temperature gradient is strongest
at low $e$, we further delete all data in each weekly map taken at $e
< 80\degree$.  The resulting maps have noise and artifacts which are
somewhat worse than the Annual Average Skymaps from the NSSDC; however,
the artifacts are the same at both wavelengths.  The resulting
$25\micron$ map is appropriate for
modelling the IPD contamination at other wavelengths.

\subsubsection{Scale and Zero-Point of the Zodiacal Light}
\label{section_zody}

The spatial-temporal variations of zodiacal light are complicated and
difficult to model analytically.
Our procedure for removing the zodiacal light
assumes that the $100\micron$ map correlates linearly with
\HI\ at high Galactic latitudes and low flux levels (\cite{boulanger96}).
Note that one should also expect some dust within the extensive ionized
$\Hplus$ zones, and that the column density ratio $N(\Hplus)/N(\HI)$
has considerable scatter in different directions (\cite{reynolds90}).
The relationship between the diffuse ionized and neutral hydrogen in any
direction is complex (\cite{reynolds95}); some of the $\Hplus$ is
associated with neutral gas, while other ionized clouds adjoin neutral
regions.  We are of course interested in measuring the dust associated
with both components, and to the degree that there is dust within
$\Hplus$ regions not associated with neutral regions, the scatter of
dust and \HI\ will be increased.  The degree of scatter between neutral
gas and dust is therefore of considerable interest, but this scatter
should not bias our procedure.

We consider the $25\micron$ map as a template of the
unwanted zodiacal foreground, as the $25\micron$ channel is so thoroughly
dominated by the IPD.  In a few regions where the Galactic cirrus
contributes more than a few per cent to the $25\micron$ emission,
we interpolate the $25\micron$ map in bins of constant ecliptic latitude.
To first order, we subtract scaled versions of the $25\micron$ map in order to
minimize the scatter between the \HI\ and zodiacal-corrected $100\micron$
and $240\micron$ emission.  We further refine this correction with the
addition of a quadratic term to account for temperature variations in the IPD.

\placetable{table_dirbebadpix}

Fitting black-body functions to the zodiacal light from $12\micron$
to $60\micron$, we find that the color temperature does not vary by
more than ten percent from ecliptic equator to the poles.
This allows us to make a first-order correction under the assumption
that the IPD is at a constant temperature.  The DIRBE $25\micron$ map,
$\RDmap_{25}$, is directly scaled to model the zodiacal contamination at
longer wavelengths.  The raw maps for these other DIRBE passbands,
$\RDmap_b$, are corrected to
\BEL{equ_zodycorr} \LDmap_{b} = \RDmap_{b} - A_{b} \RDmap_{25} - B_{b} . \EE
The coefficient $A_b$ determines the scaling
of the zodiacal emission from the $25\micron$ DIRBE passband to
passband $b$.  The zero-point offset, $B_b$, is used to account for a
multitude of non-zodiacal contributions to the $25\micron$ and
$100\micron$ maps, either from the Galaxy or extragalactic light.
Although the $25\micron$ map is
relatively free of point sources, the ten brightest sources have
been masked to their local averages (see Table \ref{table_dirbebadpix}).
Note that these sources were not excluded from the $100\micron$
or $240\micron$ maps.

\placetable{table_zodycoeff}

\begin{figure}
\plotone{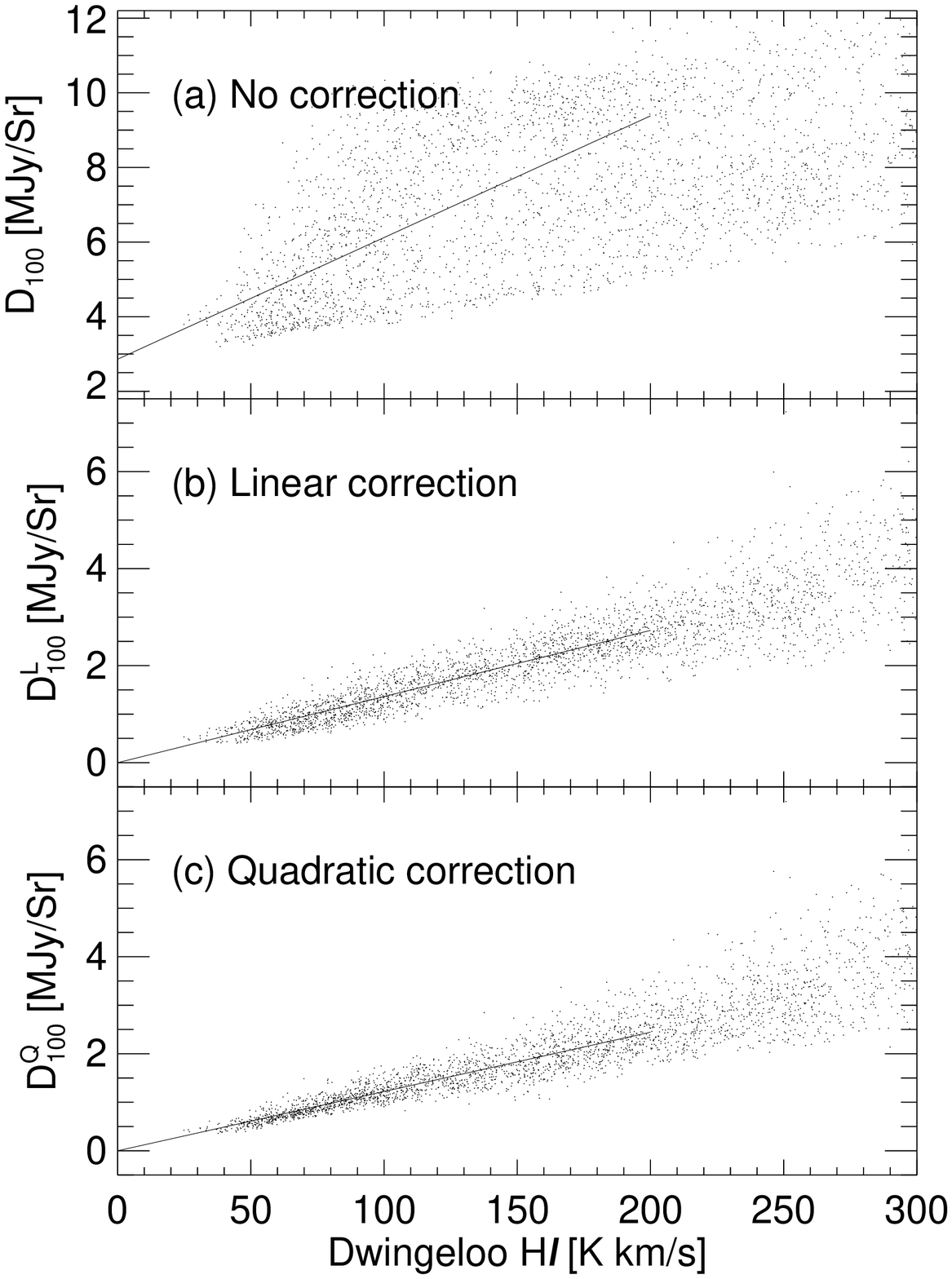}
\caption{The $100\micron$-\HI\ correlation with (a) no correction, (b) linear
  correction, and (c) quadratic correction for zodiacal contamination.
  The fits in the range $[0,200] \Kkms$ are shown as solid lines.}
\label{fig_HI100}
\end{figure}

Assuming that \HI\ correlates linearly with dust
at high latitudes and low flux levels,
we make use of the \HI\ flux to determine the zodiacal flux scaling
coefficients and zero-point offsets.   
We compare the emission, $\LDmap$, to the \HI\ column
density, $\Hmap$, from the Leiden-Dwingeloo 21-cm survey
(\cite{dwingeloo}).  The 21-cm emission is summed in the velocity
range $-72 < \vLSR < +25 \kms$, which samples almost all of the Galactic
\HI\ at high Galactic latitudes.  (At high latitudes, \HI\ outside of
this range shows no correlation with the $100\micron$ maps).  We solve
for the coefficients $A_b$ and $B_b$ of equation \ref{equ_zodycorr}
and the \HI-to-DIRBE ratio, $\zeta_b$, using a least-squares
minimization of
\BEL{equ_zodyfit} \chi^2 = \sum_i \left[ \frac{ \zeta_b (\LDmap_{b})_i
- \Hmap_i} {\sigmap_i} \right]^2 \EE
in each passband.
The errors in this relation are systematic and difficult to express
analytically.  Therefore, we take the uncertainties, $\sigmap_i$, to
be proportional to the \HI\ flux.  The fitting procedure
is performed on the maps binned to $2.5\degree$ and
limited to those areas of the sky with low emission,
\HI\ $ < 200 \Kkms$, or $N(H) < 3.7 \times 10^{18} \cm^{-2}$ ($19\%$
of the sky).  Two dust clouds with unusual temperature
and the region immediately around the very bright source
NGC~253 are omitted from the fits.
The results of these fits for $b=100$, $140$, $240\micron$ are presented in
Table \ref{table_zodycoeff}, and a scatter diagram is shown
in Figure \ref{fig_HI100} both (a) before and (b) after the correction.

\placefigure{fig_HI100}

Temperature variations in the IPD make the linear correction of equation
\ref{equ_zodycorr} inadequate for our purposes.
Based on $12\micron / 25\micron$ and $25\micron / 60\micron$ ratios,
the color temperature of the IPD varies by $5\%$ for an $\alpha=1$ emissivity
law or by $10\%$ for an $\alpha=2$ emissivity law.
Because the $100\micron$ band is on the Rayleigh-Jeans side of
the spectrum, this results in a $5$ or $10\%$ error in zodiacal light.
As the zodiacal light contributes up to $10\MJypSr$ at $100\micron$
in the ecliptic plane, this results in absolute errors of $1\MJypSr$,
with a strong ecliptic latitude dependence. 
A constant temperature, linear model of zodiacal dust removal is a good
first approximation, but is clearly too simplistic.
Because the neighboring passbands at $12\micron$ and $60\micron$ are a complex
mix of zodiacal and non-zodiacal contributions, it is desirable to model
the variation in IPD temperature using only the $25\micron$ map rather
than some linear combination of passbands.  Thus we are driven to higher order
fits using only the $25\micron$ map.

The temperature variations in the IPD are most strongly a function of
ecliptic latitude, $\beta$.  An adequate second-order function for 
modelling the variations in IPD emission at longer wavelengths is to
include a term that is linear in the product of $\RDmap_{25}$ and 
$\bar{\RDmap}_{25}(\beta)$, the mean $25\micron$ emission at
each ecliptic latitude (limited to $|b| > 20\degree$ to minimize Galactic
contamination in the model).  With a
scaling coefficient, $Q_b$, the corrected maps then become
\BEL{equ_quadcorr} \QDmap_{b} = \RDmap_{b} - 
(A_{b} + Q_{b} \bar{\RDmap}_{25}(\beta) ) \RDmap_{25} - B_{b} . \EE
The above is referred to as the quadratic correction.
Solutions are again found via equation \ref{equ_zodyfit} using the same
region of the sky as before.
Results are presented in Table \ref{table_zodycoeff}.  Since the 
linear and quadratic contributions are not orthogonal, there is no
particular physical meaning to the coefficients derived in the quadratic fits; 
they are merely the coefficients which yield the minimum $\chi^2$.

The result of the quadratic correction is shown in Figure
\ref{fig_HI100}(c).  At $100\micron$, the quadratic correction is
clearly superior to the linear correction, reducing the RMS scatter
from 19\% to 16\% (for \HI $< 200\Kkms$).
At $240\micron$, the difference between the
two fits is negligible.  This is both because the $240\micron$ map has
a higher inherent noise, and because the zodiacal light correction at
such long wavelengths is smaller.  Even though there is no obvious
difference in fitting methods at $240\micron$, we apply the quadratic
correction to both passbands to avoid introducing systematic errors
into the determination of the dust temperature.
Using the Bell Laboratories \HI\ survey (\cite{stark92}) instead
of Leiden-Dwingeloo results in a zero-point change in the final
$100\micron$ map of only $0.013\MJypSr$, with a standard deviation
of $0.047\MJypSr$.

In principle, the fit should be done between the \HI\ gas
and the total column density of dust.  The total column is a
complicated function of the $100$ and $240\micron$ maps,
and the zodiacal-light coefficients for both those maps need be fit
simultaneously.  This non-linear fit with six parameters is not stable,
primarily because the DIRBE $240\micron$ signal-to-noise ratio is too low
at high latitudes.  For this reason, we have performed the fit
independently for the $100$ and $240\micron$ maps.
As long as the temperature of the dust does not vary significantly
and systematically with ecliptic latitude, then the fit coefficients
for our zodiacal model should be valid.  As we discuss in the following
section, variations in the Galactic dust temperature at the high latitudes
used in these fits are hardly measurable above the noise.

This least-squares minimization procedure is used only to remove the
zodiacal foreground.  Our measure of the best removal is minimal scatter
between the far-IR dust emission and 21-cm emission.
There could well be an additional component of dust that is not
correlated with the \HI\ emission, such as dust associated with diffuse
$\Hplus$ regions.  In such a situation, our procedure for zodiacal
foreground removal should remain unbiased, but the interpretation of the
fit coefficients would be different.
The slope $A_b$ would signify the amount of dust associated with neutral gas
mixed with ionized gas, while the offset $B_b$ would be the dust
associated with instrumental offset, the CIB, and dust emission associated
with a uniform sphere of diffuse $\Hplus$.

This method of zodiacal light correction is by no means definitive.  A
more thorough examination of the problem is currently under way by
Reach and collaborators (\cite{reach97}).  They are performing a
detailed modelling in three-dimensional space of the IPD that includes
the dependence of dust temperature and density on the distance from
the Sun, as well as specific dust bands, and the inclination of the dust
cloud with respect to the Earth's orbit.  However, our model is adequate
for the present analysis.  In the worst case, assuming that all of the
residual scatter in the corrected $100\micron$ plots at \HI $<100 \Kkms$
is due to imperfectly-subtracted zodiacal light, the RMS of this error
is $0.05\MJypSr$.  This corresponds to an error of $0.004\MAG$ in $\Ab$.

\subsection{Zodiacal Removal to Isolate the CIB}
\label{section_cibr_fit}

As a digression at this point, the question of the cosmic infrared background
(CIB) was a major motivating factor for the DIRBE experiment.  The
extraction of the CIB has proven to be a considerable challenge
(\cite{hauser96}), and to date CIB detections from DIRBE data
have been reported
as upper limits.  The regression analysis described above is one method to
approach such a measurement, provided the zodiacal model is adequate.  The
coefficient $B$ of the regression fits is a constant term that remains after the
zodiacal foreground dust and all Galactic dust correlated with \HI\ emission is
removed.  This term is not likely to be an instrumental offset, since DIRBE had
a cold-load internal chopper and the entire telescope was cold.
A significant detection of $B$ is either a
measure of the CIB or due to some unknown instrumental artifact.

Since the zodiacal dust is confined to a thin zone in the ecliptic plane of the
solar system, along directions away from the ecliptic plane, all the zodiacal
foreground is approximately 1 A.U.  from the Sun.  It is therefore reasonable to
assume the zodiacal foreground emission of our reconstructed yearly maps to be
the same temperature in all directions, for sufficiently high $|\beta|$.  Thus
a linear model should be  adequate for removal of the zodiacal emission.
However, one cannot look strictly at the ecliptic poles, since the
$25\micron$ maps vary only a factor of 3.5 from the ecliptic plane to the pole;
a suitable lever arm in $\beta$ 
is therefore essential to decouple the isotropic CIB 
from the smoothly varying zodiacal foreground.

We have performed linear regression analyses for the $100\micron$, 
$140\micron$, and $240\micron$ maps as a function of the lower limit
on ecliptic latitude, $|\beta|$, using the procedures described in
Section \ref{section_zody}.  The results are listed in Table
\ref{table_cibr_coeff}.  The quoted errors represent the formal 95\%
confidence intervals, and do not incorporate systematic errors due
to our choice of zodiacal model.  For the zero-point term, $B$, the quoted
errors are dominated by covariance with the zodiacal coefficient, $A$.
At $100\micron$, we have no significant detection, largely because the
zodiacal foreground is quite substantial.  At $140\micron$ and
$240\micron$, the determination of $B$ is significant and non-zero.
These are possibly measurements of the CIB, as we discuss in Section
\ref{section_cibr_discuss} below.

\placetable{table_cibr_coeff}

\subsection{DIRBE Temperature Maps}
\label{section_tempmap}

The main component of Galactic dust detected by DIRBE spans
a color temperature range of $17\K$ to $21\K$
for an assumed $\lambda^{-2}$ emissivity (\cite{reach95}).
At the extremes of this temperature range, the emission at $100\micron$
will differ by a factor of $5$ for the same column density of dust.
Therefore, it is crucial to use temperature information to recover
the dust column density.
There are three useful DIRBE bands for computing the Galactic dust
temperature, centered at $100\micron$, $140\micron$ and $240\micron$.
The median signal-to-noise ratio per pixel for these maps
at $|b| > 10\degree$ is $164$, $4.2$, and $4.9$, respectively.
Because of the high $S/N$ at $100\micron$, we will use that map as a
spatial template and multiply it by a temperature-correction factor
derived from the $100\micron$ to $240\micron$ ratio.  The $140\micron$
map is not used because of its large noise level compared to the nearby
$100\micron$ map.
We attempt no temperature correction in regions of very low emission,
since our procedure will introduce noise and the exact temperature
of the dust in these dilute regions is less important.

\subsubsection{FIR Spectrum of the Dust}
\label{section_spectrum}

Estimation of the dust color temperature and column density is complicated by
our lack of knowledge about dust opacities and temperature distributions.
Our procedure works only if the large
grains responsible for the emission at $100$ to $240\micron$ are in
thermal equilibrium with the interstellar radiation field, and the
smaller, transiently heated dust grains make a negligible contribution
to emission at these wavelengths.  Fortunately, these assumptions
appear to be in agreement with our current understanding of interstellar
dust emission (\cite{guhathakurta89}; \cite{sodroski97}).

Because the ISM is optically thin in the FIR, the intensity $I_\nu$
is given by
\BE I_\nu = \int{ ds\, \rho \kappa_\nu B_\nu(T) } , \EE
where $T$ is the grain temperature, $\rho$ is the mass density,
$B_\nu$ is the Planck function, $\kappa_\nu$ is the opacity,
and the integral is over the physical path length.
Since the dust particles are small (radii $a < 0.25\micron$) compared
to FIR wavelengths, the opacity $\kappa_\nu$ does not depend upon the
details of the particle size distribution.  However, the opacity
does depend upon the nature of the material.  In general, the opacity
follows a power law in the FIR,
\BE \kappa_\nu \propto \nu^\alpha . \EE
The Draine \& Lee (1984) model predicts that the FIR emission is
predominately from graphite with $\alpha \approx 2.0$.  Layer-lattice
materials, such as amorphous carbon, have opacities $\alpha \approx
1.0$, and silicates have $\alpha \approx 1.5$.
We have chosen $\alpha=2.0$ in deriving our dust temperatures.
Though this choice
directly affects the temperature estimation, we show that our final
dust column densities are completely insensitive to this choice.  The
reason for this insensitivity is that the DIRBE passbands are
sufficiently near the Wien tail of the dust spectrum.

\begin{figure}
\plotone{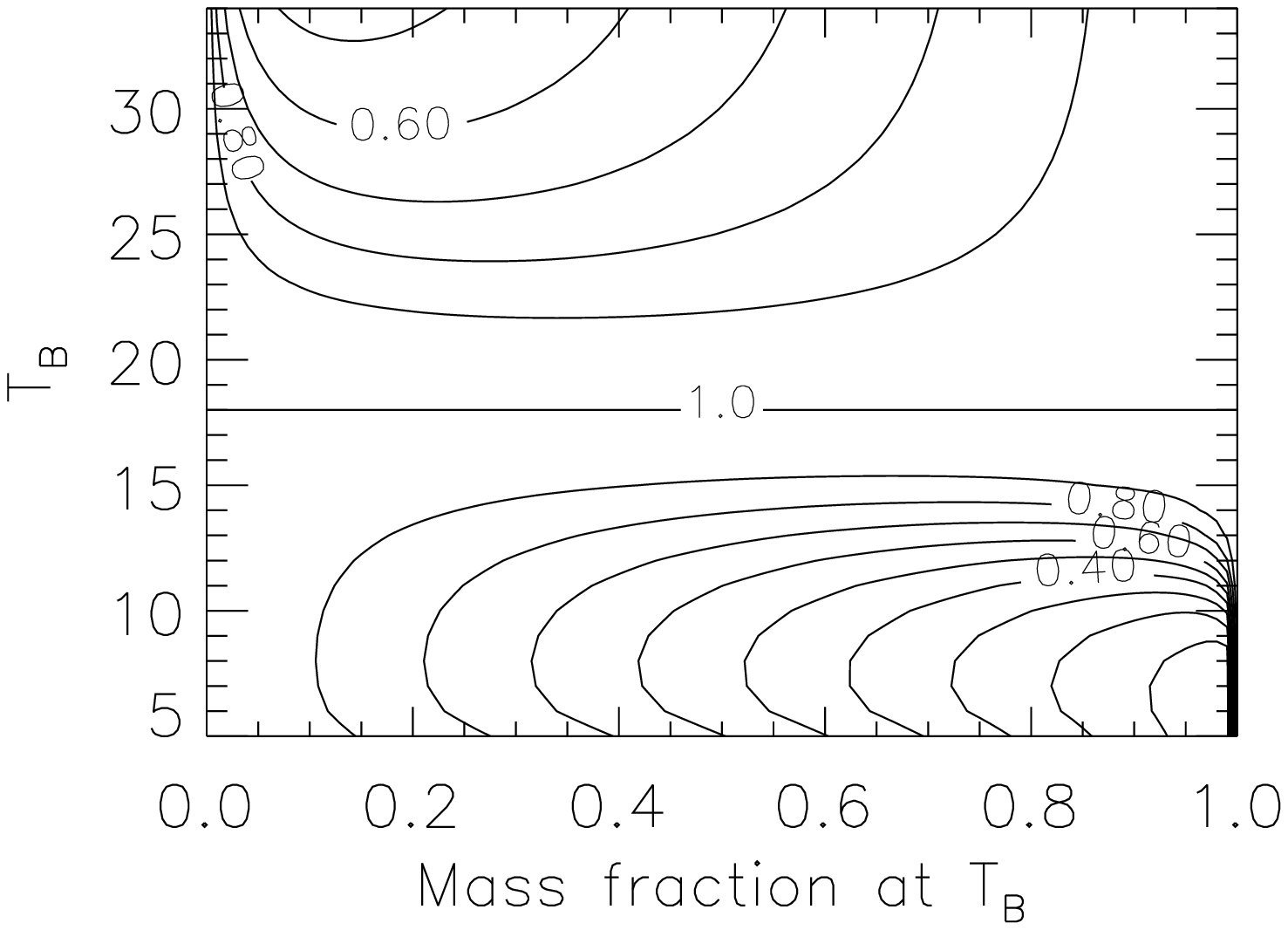}
\caption{Ratio of recovered versus true column density of dust
using a single-temperature fit to two components.
A fraction of dust, $f_B$, at temperature $T_B$ is added to $18\K$ dust.
The recovered column density is always lower than the true column density,
with contours spaced in units of $0.1$.
}
\label{fig_twotemp2}
\end{figure}

\placefigure{fig_twotemp2}

``Classical'' grains with radii $a > 0.05\micron$ exposed to ambient
starlight retain an equilibrium temperature.
In the section that follows, we assume such an equilibrium and
fit each line-of-sight with a single color temperature.
However, each line-of-sight through the Galaxy may pass through several
regions at different equilibrium temperatures.  By ignoring multiple
temperature components, we systematically underestimate the true
column density.  We have modelled this effect by fitting a single
color temperature to emission from $18\K$ dust added to emission from
another region at temperature $T_B$.  For different choices of $T_B$
and its mass fraction, $f_B$, we compare the recovered dust column
to the true column density (Figure \ref{fig_twotemp2}).
For the range $15 < T_B < 21.5\K$, the true column is at most underestimated
by $10\%$.  A factor of two error would result only in contrived
circumstances with $T_B < 12\K$ or $T_B > 33\K$.
As the majority of the sky spans only a five degree range of temperature,
our single-temperature model is deemed satisfactory.

Another constituent of the dust which we have ignored are very
small grains (VSGs), with radii $a \simlt 0.005\micron$.
These VSGs are stochastically heated by the interstellar radiation field
(ISRF), and cannot be characterized by a single temperature.
VSGs undergo a sudden rise in temperature following absorption of a visible
or UV photon, followed by a nearly-continuous decline in temperature as
the heat is radiated away in the form of many FIR photons.
Such grains are assumed to be responsible for the stronger-than-expected
Galactic cirrus emission observed by IRAS at $12- 60\micron$.
The contribution from VSGs at the wavelengths we use to fit the
temperature ($\lambda > 100\micron$) is small (\cite{draine85};
\cite{guhathakurta89}).
Thus, the VSGs do not spoil our derivation of the column density due to
classical grains.  VSGs would be significant to our dust maps only if
their density is much higher than expected, and their distribution
different from that of classical grains in the diffuse ISM.

\subsubsection{Recovering a Temperature Map}
\label{section_Trecover}

The $100\micron$ and $240\micron$ maps are too noisy to recover
a reliable and independent dust color temperature at each DIRBE pixel.
In the regions of low signal-to-noise,
a filtering scheme is required.  One option is to smooth the maps
with a large beam, using the resulting maps to construct
a heavily-smoothed temperature map.  This simple technique has the
disadvantage of throwing away small-scale information in the compact dusty
regions where the signal-to-noise ratio is high.
The spirit of most filtering methods is to produce a map with a fairly
uniform signal-to-noise ratio.
An optimal scheme would smooth the data where there is no
signal (and very little dust), and leave the data unsmoothed in the
regions with large signal (and more dust).  However, such schemes
(e.g., the minimum-variance Wiener filter)
have the disadvantage of overweighting low S/N pixels near high S/N
pixels in such a way that bright sources have noisy halos.

One way to avoid artifacts such as halos around bright sources
is to filter the maps on a local basis at each pixel.
We assume that regions of very little dust have a uniform temperature.
This allows us to combine the DIRBE maps with background averages taken
at high Galactic latitudes, weighting in such a way that we preserve S/N.

To generate the temperature-correction map from the
zodiacal-light-corrected DIRBE maps, we first remove 13 bright point
sources from the DIRBE images (including M31, the LMC, and the SMC; see
Table \ref{table_dirbebadpix}).
We smooth both the $100\micron$ and $240\micron$ maps with a Gaussian
filter of FWHM$=1.1\degree$, which determines the resolution of the
resulting temperature map.

The dust effective color temperature is recovered from the ratio
of intensities at $100\micron$ and $240\micron$.  Using the filtered DIRBE maps
from equation \ref{equ_filtmap}, we define the ratio map
\BEL{equ_colorR} \Rmap = \frac{\FDmap_{100}}{\FDmap_{240}} , \EE
where
\BEL{equ_filtmap} \FDmap_b = \Wmap \QDmap_b + (1-\Wmap) {\bar \QDmap_b} . \EE
Each filtered map, $\FDmap_b$, is a weighted average of the local
(Gaussian-smoothed) DIRBE flux and the background average flux determined at
high Galactic latitude.  The background average flux, ${\bar \QDmap_b}$,
is determined separately in the North and South Galactic hemispheres.
The weight function, $\Wmap$, determines how to filter the resulting $\Rmap$
between the local quantity and the background average.  Where $\Wmap$ is small,
the resulting $\Rmap$ will have the high latitude background value,
as appropriate in regions of low S/N.
In regions of high S/N, the weight function approaches
unity and
$\Rmap$ reduces to the local value of the flux density ratio.  The advantages
of this method are that it has no leakage from neighboring pixels and that the
scale of the smoothing is constant.  Therefore, noise generated by the ratio of
two small, uncertain numbers is suppressed.
The background averages are assessed in each hemisphere by simply averaging
the $100\micron$ and $240\micron$ flux within the window
$|b| > 75\degree$.  The ratios are $0.661$ in the north and $0.666$ in
the south.

We determine the weight function $\Wmap$ by a minimum-variance
analysis of the resulting ratio map $\Rmap$.  For this purpose we use
a map of the standard deviation of measured $240\micron$ flux at each
point, $\sigmap_{240}$, reduced by the effective binning of the
Gaussian smoothing of the maps.  For the $240\micron$ map, the DIRBE
errors are in the range $0.5 < \sigmap_{240} < 2.0\MJypSr$, much larger
than for the $100\micron$ map.  Because the detector noise is negligible for
the $100\micron$ data, we construct a weight function that
minimizes the variance in the reciprocal of the color ratio,
$\delta^2 (1/\Rmap)$,
\BEL{equ-variance}
\left<\delta^2 (1/\Rmap)\right> = 
{{ \sigmap_{240}^2 \Wmap^2 + 
\left( {\bar\QDmap_{240}} -{\bar\QDmap_{100}}/R_T\right)^2 
\left(1-\Wmap\right)^2 }
\over {\left(\QDmap_{100} - {\bar\QDmap_{100}}\right)^2\Wmap^2 
+ \left({\bar\QDmap_{100}}\right)^2 \left(1-\Wmap\right)^2 }}
 \EE
where $R_T$ is an assumed mean color ratio.  We use $R_T=0.55$, but the 
results are insensitive to this choice.
Minimization of this function with respect to $\Wmap$ leads to a quadratic
equation whose solution is a map of the weight function
\BEL{equ_weight}
 \Wmap = \Wmap(\QDmap_{100},R_T,{\bar\QDmap_{240}},{\bar\QDmap_{100}},
 \sigmap_{240}) . \EE
The resulting $\Wmap$ function is
close to the simple expression $\Wmap = \QDmap_{240}/(\QDmap_{240} + 5)$ for
typical noise levels in the DIRBE $240\micron$ map.

\placefigure{plate_Tmap}

\placetable{table_kfit}

To convert from the ratio map, $\Rmap$, to effective color temperature,
we need to know the emissivity model for the dust, $\epsilon_\nu$,
and the frequency response of the DIRBE instrument passbands, $W(\nu)$.
This information is combined to form a color-correction factor, $K_b$,
for a passband $b$.
Using an $\epsilon_\nu = \nu^\alpha$ emissivity model,
\BE K_b (\alpha,T) = \int{ d\nu B_\nu(T) \nu^\alpha W_b(\nu) } , \EE
%
The DIRBE Explanatory Supplement (1995), Appendix B,
has tabulated $K_b(\alpha,T)$ for all their bands in the domain
$0 < \alpha \le 2$ and $10 \K \le T \le 2\times 10^4 \K$.
These color-correction factors are well fit by a functional form
\BEL{equ_KofT} K(\alpha,T) = {\sum_i a_i \tau^i \over \sum_j b_j \tau^j } ;
 \qquad \tau \equiv \LOGTEN{T} . \EE
Coefficients for these fits at $\alpha=2.0$ are presented in
Table \ref{table_kfit}.  The fits are accurate to one percent at all
temperatures.
The measured flux in passband $b$, ${\bf D}_b$,
of dust emitting thermally with a $\nu^\alpha$
emissivity is
\BE \RDmap_b = K_b (\alpha,T) I_b(T) , \EE
where $I_b$ is the actual intensity at the frequency $b$.
Calculating the ratio, $R(\alpha,T)$, that DIRBE would measure for a given
temperature and emissivity model,
\BEL{equ_RofT} \Rmap(\alpha,T) = { \RDmap_{100} \over \RDmap_{240} } =
 { K_{100}(\alpha,T) I_{100}(T) \over K_{240}(\alpha,T) I_{240}(T) } . \EE
We interpolate the function $K_b(\alpha,T)$ for 3000 temperatures and
invert equation \ref{equ_RofT}, yielding $T(R)$.  The emissivity model
$\alpha=1.5$ yields a color temperature $\sim 2 \K$ hotter than
$\alpha=2.0$, but the effect of this on the recovered dust column
density is not significant aside from an overall, multiplicative
normalization.  Aside from this normalization, the final column
density maps vary only at the $1\%$ level for an $\alpha=1.5$ versus
$\alpha=2.0$ emissivity model.  For all calculations that follow, we
have chosen $\alpha=2.0$ in agreement with the Draine \& Lee (1984)
dust model.  The resulting temperature map is shown as Figure
\ref{plate_Tmap}.  
The light regions on the temperature map denote warmer dust, while the darker
regions denote cooler dust where the $100\micron$ flux underestimates the
column density of dust.  Note that most of the sky is a neutral grey
color, while several prominent, high latitude molecular cloud regions
show up as dark filaments.  A few isolated white (hot) spots correspond to the
LMC, SMC, and O-star regions near Ophiuchus and Orion.

We compute the column density of dust as the amount of $100 \micron$ emission
we would expect if the dust were all 
at a reference temperature of $T_0 = 18.2 \K$.
The column density, $\NDmap$, is expressed as the $100\micron$ flux
multiplied by a temperature-correction factor, $\Xmap$:
\BEL{equ_NDmap_dirbe} \NDmap = \QDmap_{100} \Xmap , \EE
where
\BEL{equ_XofT} \Xmap(\alpha,T) =
  { B(T_0) \Kmap_{100} (\alpha,T_0) \over B(T) \Kmap_{100} (\alpha,T) } \EE
and $B(T)$ is the Planck function.
For an $\alpha=2.0$ emissivity model and
an arbitrary reference temperature of $T_0=18.2\K$, we combine equations
\ref{equ_RofT} and \ref{equ_XofT} to solve for $\Xmap$ as a function of the
DIRBE $100\micron / 240\micron$ ratio.  The solution is well fit by
\BE \LOGTEN \Xmap = -0.28806 - 1.85050~(\LOGTEN \Rmap)
 - 0.02155~(\LOGTEN \Rmap)^2 . \EE
A temperature map, $\Tmap$, is recovered with the function,
\BE \LOGTEN \Tmap = 1.30274 + 0.26266~(\LOGTEN \Rmap)
 + 0.04935~(\LOGTEN \Rmap)^2 . \EE
The maximum deviation of these fits is $0.8\%$ in the domain $10 < T < 32 \K$.
The column density map, $\NDmap$, recovered from equation \ref{equ_NDmap_dirbe},
is our model for the Galactic dust at DIRBE resolution.  


\section{Folding in IRAS Resolution}
\label{section_iras}

The infrared maps from the COBE satellite have the advantages of being
well-calibrated and well-corrected for zero point drift.  However,
with a resolution of $0.\degree 7$ FWHM, the COBE/DIRBE maps utterly
fail to resolve important filamentary details in the cirrus and
detect only the very brightest point sources.  Therefore, we resort
to use of the older IRAS satellite data to improve upon this limited
angular resolution.

The IRAS satellite observed $97\%$ of the sky in four passbands,
detecting some $250,000$ point sources, including $\sim 20,000$
galaxies.  In addition to these point sources, the satellite
discovered the infrared cirrus, especially visible in the
$100\micron$ band (\cite{low84}).  As the instrument was designed for
point source detection (differential photometry), it is far from
optimal for the absolute photometry required to make sense of the
cirrus.  In particular, the zero-point from one scan to the next
drifts considerably, leaving stripes with power on a range of scales,
from a few arcminutes to several degrees.

The IRAS Sky Survey Atlas (ISSA: \cite{issa94}) released by IPAC in
1991-1994 is the most useful presentation of the IRAS data for our
purposes.  The ISSA sky is divided into 430 plates 12.5 degrees square
and spaced roughly every 10 degrees in right ascension and declination.
The ISSA maps suffer from the zero-point drifts as well as residual zodiacal
light.  Despite the best efforts at IPAC to remove striping artifacts and
zodiacal light, these features are still quite obviously present.
This limits the usefulness of these maps for many applications.

We have reprocessed the ISSA data to create a $100\micron$ map 
with the following properties:
\begin{itemize}
\item ISSA plates destriped, reducing the amplitude of the
      striping artifacts by a factor of $\sim 10$,
\item de-glitching algorithm employed to remove small-scale artifacts
      that are not confirmed on other IRAS scans,
\item IRAS missing data areas filled in with DIRBE data,
\item IRAS zero-point drifts corrected to DIRBE on scales $\geq 1\degree$,
\item angular resolution close to the IRAS limit (FWHM$=6.1\arcmin$),
\item extragalactic objects removed to a flux limit of $f_{100}\approx
      1.2\Jy$ at $|b| > 5\degree$,
\item stars removed to a flux limit of $f_{100}\approx 0.3\Jy$ at $|b|
      > 5\degree$.
\end{itemize}
The combination of our destriping algorithm and the DIRBE zero-point
in effect removes zodiacal contamination from the ISSA maps.
Section \ref{section_destripe} discusses our Fourier destriping algorithm.
Removal of deviant pixels, including asteroids, is described in
Section \ref{section_deglitch}.
Section \ref{section_combine} describes the method used to combine DIRBE
and IRAS data to form our final maps.  The removal of $20,000$ point
sources is the topic of Section \ref{section_sources}.

\subsection{De-striping the ISSA Maps}
\label{section_destripe}

IPAC took steps to destripe the ISSA plates, including a zodiacal
light model and local destriping algorithm.  However, serious striping
remains.  Fortunately, the scanning strategy of the IRAS satellite gives
sufficient information to remove most of the striping artifacts that remain.
Cao \etal\ (1996) have made important progress towards destriping
the maps and deconvolving the IRAS beam reponse in small regions of
the sky.  However, due to the complexity of their
algorithm and large amount of computing time required, it is
impractical at this time to apply such a method to the full sky.  We
have developed an algorithm that is straightforward to implement and
can easily be applied to the full sky.

\placefigure{plate_fourier}

We consider each ISSA plate separately, and manipulate it in Fourier
space.  One ISSA plate and its Fourier transform are seen in Figure
\ref{plate_fourier}(a) and (b), where zero wavenumber is at the center.
The rays emanating from the center are a clear signature of the striping in
real space.  The problem then becomes one of removing this excess
power in Fourier space that is associated with the real-space stripes.
Because the IRAS satellite scanned most of the sky two or three times,
usually at a different angle, the contamination in Fourier space
occurs at different wavenumbers for each scan.  Thus, the contaminated
wavenumbers from one scanning angle can be replaced with wavenumbers
from the other scanning angles.  The steps involved are (1)
identifying the contaminated wavenumbers, and (2) an optimal
replacement strategy which allows for the lack of full coverage for
many of the scans.

Because the stripes are radial in Fourier space, we parameterize the
Fourier domain in polar coordinates $(k_r, k_\theta)$ for the purposes
of this discussion, although it is discretely sampled in $k_x$ and
$k_y$ for calculations.
 
For each ISSA plate, there are 3 HCONs (hours-confirmed scans)
referred to as HCON-1, HCON-2, and HCON-3.  The composite map generated
from these is called HCON-0, but when we refer to ``the HCONs'' we mean
HCONs 1, 2, and 3.  Many of these individual HCONs contain data with
striping in only one direction.  In general, however, a single HCON
image may contain data taken in multiple scan directions, and
significant regions of missing data.  In many cases, HCON-3
contains little or no data.  For each HCON map there is a
corresponding coverage map, which was used by IPAC to compute a
weighted average of the three HCONs for each plate.  (These coverage
maps were kindly provided to us by S.\ Wheelock).  The Explanatory
Supplement also reports the use of a local destriper, but it appears
to have had negligible effect in low-flux regions of the sky.

\subsubsection{Criterion for Bad Wavenumbers}

With each plate, we begin by generating a composite map from the three
HCONs, using the three coverage maps for weights.  This is equivalent
to the HCON-0 composite maps released with the ISSA, except that no
local destriper has been applied, and asteroids and other artifacts
have not been removed.

Because our method is based on Fourier transforms where bright point
sources may overwhelm many modes, we perform a crude point source
removal before any analysis.  This consists of selecting objects or
regions which are more than $0.7\MJypSr$ above the median-filtered
image.  Several tests of this method indicate that it masks all
significant point sources but not the striping.
These sources are then replaced with the median of an annulus around them.
Because these sources are only removed when identifying contaminated modes,
the details of this process are unimportant.

This composite HCON-0 image (without point sources) is smoothed with a
$6\arcmin$-FWHM Gaussian to produce a baseline image which is used to
fill in missing pixels in each of the three HCONs.  This smoothing is
enough to suppress most of the striping, and the replacement of missing
pixels allows us to Fourier transform completely-filled, square images.
Before transforming, we subtract the mean value
from each image, apodize the edges, and embed the $500\times 500$
image in a $1024\times 1024$ array (which is equivalent to demanding
isolated boundary conditions).
The transform for a typical plate is shown in Figure \ref{plate_fourier}(b).

It is necessary to define a criterion with which to select the
contaminated wavenumbers.  We denote the FFT of HCON-$n$ as $F_n$ and
a particular element of $k$-space as $F_n(k_r,k_\theta)$.  We bin
the Fourier domain into 90 azimuthal bins, each 2 degrees wide in $k_\theta$.
A background power as a function of $k_\theta$ is determined by first
finding the median at each $\theta$, then median-filtering the result.
We define a power ratio, $\gamma_\theta$, as the ratio of the power
in each $k_\theta$ bin to the background power.
By normalizing in this way, we expect
$\gamma_\theta$ to be near unity except in angular bins that are
contaminated by stripes.
We somewhat arbitrarily choose $\gamma_\theta < 1.2$ to be ``good''
power and $\gamma_\theta > 1.6$ to be ``bad.''
Between these two cuts, the information is deemed acceptable for
retention, but poor enough to avoid replacing other ``bad'' power with it.
In this way, ``good'' power is used where available,
and we avoid corrupting real structures by not discarding power whose
ratio is less than $1.6$.

\subsubsection{The Destriping Algorithm}

Each HCON is destriped separately before combining them to an averaged map.
Because of the incomplete coverage in some HCONs, it would not be optimal
to combine the images in Fourier space.
Instead, for each HCON-$n$, we create a stripe image, $S_n$, which is
removed in real space.
In this way, we may utilize the good data in each HCON, even if only
a fraction of the plate is covered.

A composite of ``good'' power, $F_{good}$, is generated by averaging
all power at each $(k_r,k_\theta)$ with $\gamma_\theta < 1.2$.
Any mode with no ``good'' power is set to zero.
(Modes with wavelengths larger than $1\degree$ are not changed to avoid
discreteness problems.)
A bad pixel mask is produced for each HCON, indicating which regions of
each $F_n$ are contaminated.  The difference
between $F_n$ and $F_{good}$ for bad modes only is inverse-transformed to
obtain $S_n$, the stripes corresponding to HCON-$n$.  This stripe map
is subtracted from each raw HCON to produce individual, destriped HCONs.
Note that only stripes have been removed, and the destriped images
retain all of the original point sources and other real structures.
These destriped images are then de-glitched (following section) and averaged,
weighted by the coverage maps, to obtain the final HCON-4 images.

\begin{figure}
\plotone{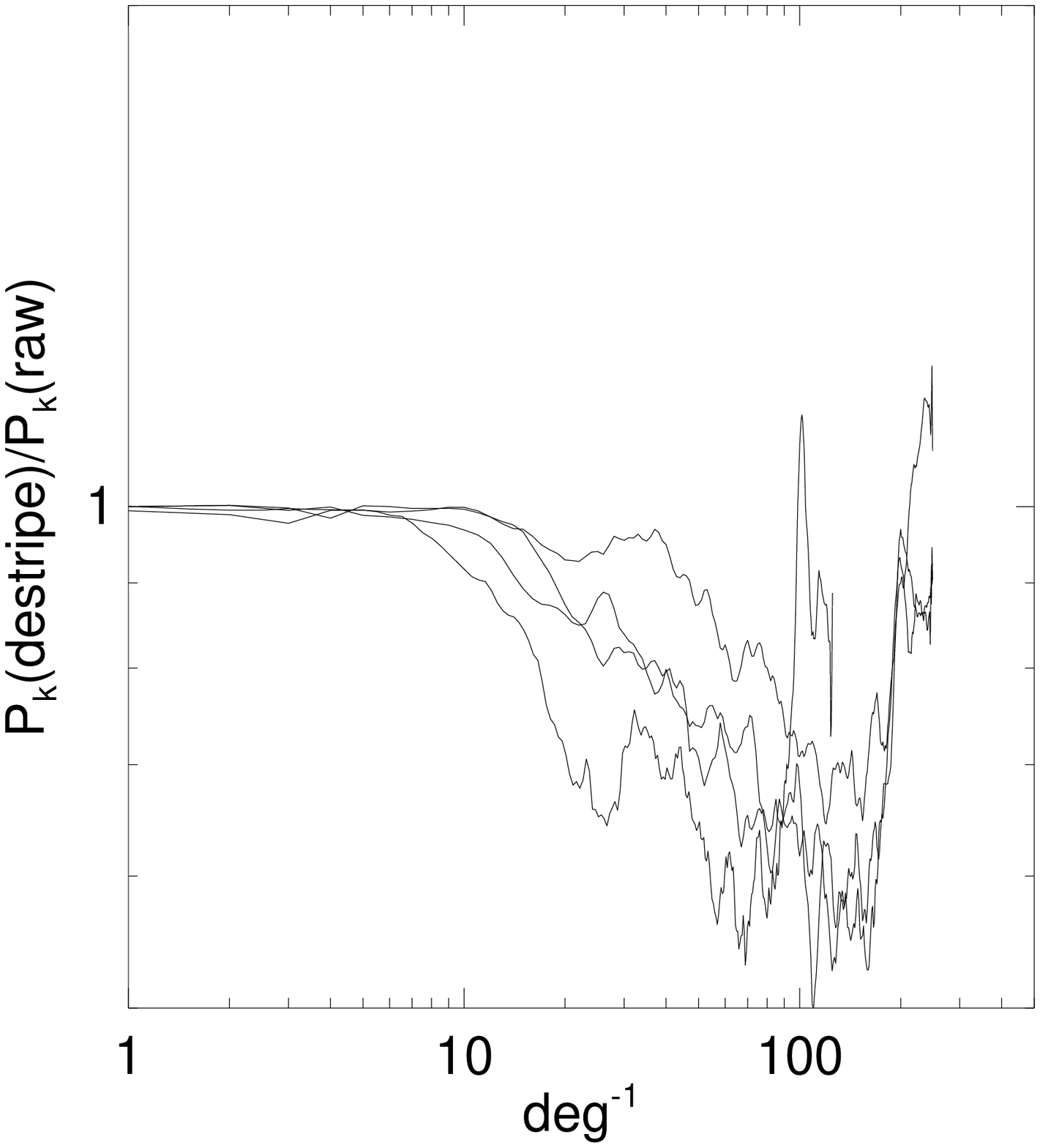}
\caption{Ratio of power in destriped ISSA image versus power in raw image.
  Ratios for four arbitrary plates are shown (plate numbers 199, 216, 217
  and 379).}
\label{fig_pkratio}
\end{figure}

\placefigure{fig_pkratio}

Figure \ref{fig_pkratio} shows the one-dimensional power spectrum
of four plates before and after the destriping process.  The
destriping has reduced the high spatial-frequency variance by nearly a
factor of two in some high Galactic-latitude areas.
In cases where the point sources subtraction fails, nothing is done to
the plate.  This occurs only at low Galactic latitude where
the striping artifacts are overwhelmed by cirrus emission.
We have not found stripes to be important for the generation of
a reddening map in these areas, so we have done nothing about them.

\subsection{De-glitching the ISSA Maps}
\label{section_deglitch}

In addition to striping, the ISSA images contain other artifacts that
must be removed.  These anomalous features include transient sources,
such as asteroids, and detector glitches that should have been removed
from the time-ordered data stream.
Images with two or more HCONs allow identification of these artifacts
by looking for discrepant pixels.
We generically term this process ``de-glitching.''
IPAC chose to de-glitch the ISSA images by visual inspection.
This post-production step removed anomalies from individual HCONs before
combining them to produce HCON-0 images.  Thus, HCON-0 differs from
a coverage-map-weighted average of the HCONs wherever anomalies were removed.
Rather than rely upon visual inspection, we implement an automated
de-glitching algorithm.

The first task is to identify discrepant points.  We compare each
destriped HCON to a local median, filtered on a $10.\arcmin$ scale.  Because
some ISSA pixels have better coverage than others, they have lower
noise.  Therefore, we set the threshold for a deviant pixel at
$2/\sqrt{N}\MJypSr$, where $N$ is the number of times IRAS observed
that pixel.  The pixels which deviate from the
median-filtered background by more than this threshold are flagged,
as are their immediate neighbors.  Pixels flagged in all available
HCONs are deemed to be ``confirmed sources.''  These include real
point sources, knots in the cirrus, and also pixels near regions of no
data.  In all these regions, asteroids cannot be differentiated
from features that should remain in the map.  However, pixels that are
flagged in at least one, but not all, HCONs are flagged as ``glitches.''
These pixels and their neighbors are replaced with data from the other HCONs.
The reader should note that cirrus features are not modified by this
algorithm.

Our de-glitching algorithm affects less than $0.1$ percent of the sky.
For comparison, IPAC removes $0.47\%$ of pixels at $|\beta| < 20\degree$
and $0.19\%$ of pixels at $|\beta| > 20\degree$.
The resulting destriped and de-glitched ISSA maps are termed HCON-4.

Three planets were scanned by IRAS and not adequately removed
from the ISSA plates.  These require special attention, as Saturn in particular
is enormously bright compared to the background flux.
Neptune and Uranus moved sufficiently between scans that they can be
removed in one HCON while retaining coverage in the other HCONs.
Because Neptune appears close to the Galactic plane,
its surface brightness is comparable to the background level of
$\sim 100 \MJypSr$.  We remove it in each HCON with a circle of radius
$7.5\arcmin$.
Uranus is only incompletely removed from the ISSA plates, still causing
hysteresis in the in-scan direction.  We remove areas of $220\arcmin$
in the in-scan direction and $94\arcmin$ in the cross-scan.
Saturn is problematic since it appears only in HCON-1 and HCON-2,
and moves very little between those scans.  Because of this, we
excise a circle of radius $2\degree$ in all HCONs.  This area,
centered at $(l,b)=(326.28\degree,+51.66\degree)$, is treated like the
exclusion strip and filled with DIRBE data (see following section).
The planet removal is implemented before the destriping algorithm,
and the de-glitching procedure afterwards.

\subsection{Combining ISSA with DIRBE}
\label{section_combine}

The ISSA and DIRBE data are combined in such a way as to retain
the small-scale information from IRAS and the large-scale calibration of DIRBE.
The ISSA HCON-4 maps are processed to match the color response and zero-point
of the zodiacal-subtracted DIRBE $100\micron$ maps on scales larger than one 
degree.  Regions with no IRAS data are filled with DIRBE-resolution data.
The maps are first projected into two $4096\times 4096$ 
pixel polar projections, one for the north and another for the south.

The ISSA maps are first multiplied by a constant value, $C$, to approximately
correct the IRAS gain to that of DIRBE.
The relative gain factors between DIRBE and IRAS are difficult to
assess exactly, owing to the drifts in IRAS zero-points.
The DIRBE Explanatory Supplement presents a preliminary linear
transformation between IRAS and DIRBE data based upon carefully selected
regions (in an obsolete version of the Supplement, but reprinted in
\cite{issa94}, Table IV.D.1).
They find the IRAS brightness levels to be too high by $38\%$
at $100\micron$.  This is a compromise value, as we find the IRAS calibration
to be too high by this value for the cirrus, yet consistent with DIRBE
for bright point sources.  We choose to globally recalibrate
the IRAS data with a compromise value of $C=0.87$.
We also convolve the ISSA images with a FWHM$=3.2\arcmin$ Gaussian,
$\Wg (3.2\arcmin)$, bringing the effective IRAS smoothing to $6.1\arcmin$.

In order to correct the IRAS zero-point to that of DIRBE on scales $>1\degree$,
we construct a difference map between the two on these scales.
This difference map, $\Diffmap$, is added to the destriped IRAS map, $\Ides$,
to yield a final map containing the small-scale information from IRAS with
the large-scale calibration of DIRBE (after zodiacal correction):
\BE \Icorr = C \cdot \Ides * \Wg (3.2\arcmin) + \Diffmap . \EE
The difference map must be taken between the IRAS and DIRBE data filtered
to the same point spread function.  The procedure is complicated by the fact
that neither the IRAS nor DIRBE beams are Gaussian.
The DIRBE point spread function is approximately $42\arcmin$ square
with power-law tails.  Fortunately, for most of
the sky, repeated scans from different directions result in an averaged DIRBE
response approximated by a circular tophat.  Therefore, the ISSA
map is convolved with a circular tophat of radius $21\arcmin$, $\Wsq$,
while still in the $1.5\arcmin$ pixel plates, then is re-projected
to a polar projection,
introducing the same distortions that exist in the DIRBE projections.
At this point, both maps are further smoothed by a FWHM$=40.\arcmin$ Gaussian
to obtain a final map with FWHM$=1.00\degree$.  The difference map is
\BE \Diffmap = [ \QDmap - C \cdot \Ides * \Wg (3.2\arcmin) * \Wsq ]
 * \Wg (40\arcmin) , \EE
where $\QDmap$ represents the $100\micron$ DIRBE map after zodiacal correction
(equation \ref{equ_quadcorr}).  The RMS of the difference map is $0.9\MJypSr$
at $|b| > 30\degree$, which is representative of the zero-point drifts in
the IRAS/ISSA maps.

Due to the differing color responses, the residual of the smoothed
ISSA and DIRBE maps has a few high and low regions, near very bright
sources.  Other than NGC~253, these are all in the Galactic plane, the
LMC or the SMC.  We mask these regions in the difference map to the
median of an annulus around them (see Table \ref{table_dirbebadpix}).
This treatment is not meant to be
strictly correct, and introduces substantial absolute errors, relative
to the rest of the map.  However, on top of the sources, the
fractional errors are not significant, and this procedure eliminates halos
and recovers the correct zero-point nearby the sources.

Regions of missing ISSA data must be filled with DIRBE data in a way
that treats the boundaries of the missing ISSA data properly.  Both steps
of the ISSA smoothing process are done in the following way: a mask,
set to 1 for good pixels and 0 for missing data, is smoothed in
the same way as the data.  The missing data is zero-filled so that
it does not contribute to the smoothed result.  After smoothing,
mask pixels with values less than $0.5$ are discarded, and the
remaining pixels are properly weighted by dividing
by the value in the smoothed mask.
The mask generated by the second ISSA smoothing is applied to the
DIRBE data set as well, so that in both maps, pixels near a boundary
only contain information from the ``good'' side of the boundary.
Point sources just barely on the ``bad'' side should contribute
equal tails to the ``good'' side, eliminating potential problems
with the zero calibration.

The column density of dust that radiates at $100-240\micron$ is recovered with
\BEL{equ_NDmap_iras} \NDmap = \Icorr \Xmap , \EE
where we have replaced the DIRBE $100\micron$ flux with the corrected
IRAS $100\micron$ flux in equation \ref{equ_NDmap_dirbe}.
The temperature-correction
map, $\Xmap$, is necessarily only at the DIRBE resolution of $\sim 1\degree$,
although the reprocessed IRAS map has a final resolution of $6.1\arcmin$.


\section{Removing Point Source and Extragalactic Objects}
\label{section_sources}

For the purposes of an extinction map, only the Galactic infrared cirrus is
of interest.  Our dust maps are meant to trace the diffuse emission from dust
as well as localized clumps, such as Bok
globules, which are typically on the scale of a few arcminutes.
The stars, planetary nebulae and other
point (unresolved) sources in the ISSA must be eliminated.
We also remove verified extragalactic objects down to a given flux limit.
The resulting maps should be free of all contaminating sources and
galaxies at $|b| > 5 \degree$, and most such sources in select regions
at lower latitudes.

\subsection{Strategy}

Our goal is to replace contaminating sources with the most likely
value of the underlying $100\micron$ emission.  The most obvious
solution is to subtract the flux multiplied by the PSF for each source.
However, the shape of the PSF is neither Gaussian nor circularly symmetric,
and its shape depends upon the IRAS scan directions at each position on the sky.
Because of these uncertainties, and because many extragalactic objects 
are not point sources anyway, we replace sources with a median value
from the surrounding sky.  For point sources, we replace pixels within
a radius of either $5.25\arcmin$ (for $f_{100} < 10 \Jy$)
or $7.5\arcmin$ (for $f_{100} \ge 10 \Jy$).
For extended sources, we replace pixels
within the measured radius plus $1.5\arcmin$.  The median is always taken from
a surrounding annulus with the same area as its interior.
In total, $1.2\%$ of the ISSA pixels at $|b| > 5\degree$
are flagged as sources and replaced.
The LMC, SMC and M31 and sources in their vicinity are not removed.
At latitudes $|b| < 5\degree$, we remove galaxies and stars only in those parts
of the sky deemed unconfused by the PSCZ galaxy survey (\cite{saunders97}).

Sources are not identified from the IRAS $100\micron$ maps due to the
confusion with cirrus.  Instead, sources are primarily identified
from the $60\micron$ band of the IRAS Point Source Catalog (\cite{psc88}),
where confusion with cirrus is less of a problem and the spatial resolution is
somewhat better.

\subsection{Extragalactic Sources}

Galaxies detected by the IRAS satellite have been extensively studied,
and lists of such objects are readily available from the literature.
Rice \etal\ (1988) have studied the largest galaxies detected by IRAS.
Strauss \etal\ (1992) and Fisher \etal\ (1995) completed a redshift
survey of IRAS galaxies to a flux limit of $1.2\Jy$ at $60\micron$
over most of the sky.  The PSCZ redshift survey (\cite{saunders97}) 
covers a somewhat smaller area of sky to a lower flux limit of $0.6\Jy$.

Many nearby galaxies are resolved in the ISSA maps and cannot be
treated as point sources.  Rice \etal\ have studied those galaxies
with blue-light isophotal diameters greater than $8\arcmin$,
and reports improved total flux densities from the IRAS data.
The 70 large galaxies with a total flux at $60\micron$ exceeding $0.6\Jy$
are removed from our maps as extended objects (excepting the LMC,
SMC, M31 and M32, due to its proximity to M31).
These objects are removed using a radius equal to one-half their major-axis,
isophotal diameter at $\Bband=25\magpasec$, plus $1.5\arcmin$.
The radius of NGC~253 is increased to $36\arcmin$ to remove IRAS hysteresis
artifacts near this extremely bright source.

The IRAS  1.2 Jy Galaxy Survey identifies most infrared-emitting
galaxies at $|b| > 5\degree$ down to a flux limit of $1.2\Jy$ at $60\micron$
(\cite{fisher95}).  Fluxes for extended sources were systematically
underestimated by the PSC, and were improved with a process called
ADDSCANing (\cite{strauss90a}).
The galaxy survey employed a color selection,
$ f_{60}^2 > f_{12}f_{25}$, which very efficiently discriminates
between galaxies and stars.  Each candidate source was classified
based upon visual inspection of POSS and ESO plates, or upon inspection
of CCD images if the plates were inconclusive.
Planetary nebulae were identified spectroscopically.
The survey identifies 5320 galaxies with $f_{60} > 1.2 \Jy$.
In addition, 444 point sources made the flux and color cuts:
98 extragalactic \HII\ regions, 210 stars, and 136 planetary nebulae.
All these galaxies and point sources
are removed from our maps, unless they coincide with the large (Rice \etal)
galaxies already removed, or the LMC, SMC and M31.
We ignore the sources classified as ``cirrus or dark cloud'',
``unobserved'' or ``empty'' fields, and two reflection nebulae.

The PSCZ redshift survey covers nearly the same region of sky as the
1.2 Jy survey, but to a flux limit of $0.6\Jy$ at $60\micron$.
This survey employed an additional color cut, $f_{100}/f_{60} < 4$,
which retains approximately $97\%$ of galaxies.
A list of PSCZ identifications was provided by Will Saunders (1997)
before publication.
This list includes $15,285$ galaxies, many of which are duplicates
from the 1.2 Jy Survey.  These galaxies, plus an additional $455$ objects
measured as unresolved by the ADDSCAN process, are removed from the
dust maps.

The 1.2 Jy and PSCZ surveys did not have complete sky coverage.
Two strips in ecliptic longitude (the exclusion strip)
were not observed by the IRAS satellite,
amounting to $4\%$ of the sky.  Other regions, primarily at low Galactic
latitude, are too confused by cirrus emission to make reliable source
identifications.  These galaxy surveys divided the sky into approximately
one-degree-square ``lune'' bins, from which masks were constructed.
The 1.2 Jy survey masks those bins at $|b| < 5 \degree$, in the exclusion
strip, or with more than 16 sources per square degree.  These ``confused''
bins comprise $3.5\%$ of the sky at $|b|>5\degree$, and are primarily
located in star-forming regions such as Orion-Taurus,
Ophiuchus and Magellanic Clouds.
The PSCZ survey masks a larger fraction of the sky at high latitudes
($8.5\%$), but includes $1.2\%$ of the sky at low latitudes which is
not confused.

We identify and remove a small number of galaxies by hand to
make our dust maps as free as possible from contamination at $|b| > 5\degree$.
The Point Source Catalog was searched for objects brighter than $1.2\Jy$
that were masked from both the 1.2 Jy and PSCZ surveys.  $195$ satisfied
the PSCZ color criteria for galaxies.  Upon visual inspection of ISSA plates
and the Digitized Sky Survey, $68$ were identified as point sources
or galaxies, and removed from the dust maps.

The many faint galaxies that remain in our maps are not significant.
As the median galaxy color
is $f_{100}/f_{60}=2.0$, our adopted flux cut at $60\micron$ roughly
corresponds to $f_{100} > 1.2\Jy$.  Galaxies just below this limit
will have a peak flux density of $\sim 0.12 \MJypSr$ at $100\micron$,
corresponding to $\Ab \approx 0.01 \MAG$.
Thus, the final extinction maps will have some small-scale
contamination from extragalactic objects at that level.
This contamination is very nearly uniformly distributed.
In addition, very faint galaxies must contribute at some level 
to a uniform,
extragalactic flux.  This contribution cannot be easily extrapolated from
known galaxies, since the measured number counts,
$N \propto f_{60}^{-1.4} $, must flatten at low flux levels to prevent
the total number and total flux from diverging.
However, any uniform contamination is irrelevant, as it is degenerate
with the constant term in the
zodiacal contamination which has already been removed
(Section \ref{section_zody}).

\subsection{Stars}
\label{section_stars}

We remove stars from the ISSA maps in all regions of $|b| > 5 \degree$,
and select regions of lower latitude.  Strauss \etal\ (1990) found that
the infrared color criterion
\BE f_{60}^2 < f_{12} f_{25} \EE
effectively discriminates stars from galaxies.
We identify as stars all PSC objects that satisfy this color cut.
At latitudes $|b| > 5 \degree$ (and not near M31, LMC and SMC), 
$4697$ sources with $f_{60} > 0.6 \Jy$ are selected for removal from
the dust maps.
At lower latitudes, we only select stars in the unmasked regions
of the PSCZ survey.  A higher flux cut of $f_{60} > 1.2 \Jy$ selects
$97$ stars in those low-latitude zones.

In Appendix \ref{section_contam}, we show that any remaining, fainter stars
produce negligible contamination.


\section{Using the Maps to Measure Galactic Reddening}
\label{section_calibrate}

For the past 17 years, almost all extragalactic observers
have made reddening and extinction corrections using the
maps of Burstein \& Heiles (BH: 1978, 1982).
The BH maps derive the column density of dust from \HI\ 21-cm flux.
At northern declinations ($\delta > -23\degree$), the BH maps
are modulated by local \HI-to-dust ratios on scales of $13~{\rm deg}^2$,
derived from the Shane-Wirtanen galaxy counts.
The DIRBE/IRAS dust maps, on the other hand, directly measure the dust,
and no corrections
need be made for optically thick \HI\ emission, for ionization, or for the
formation of molecular hydrogen.  Furthermore, the DIRBE/IRAS maps are
of uniform quality over the full sky, whereas the BH maps exclude
all regions at $|b| < 10\degree$
($17.4 \%$ of the sky), as well as a region of $1080$ square degrees
($2.6 \%$ of the sky) lacking 21-cm data.  This latter region is
near the south Galactic pole, approximately bounded by
$-135\degree < l < 21\degree$ and $b < -62\degree$.

We limit the discussion of tests of our reddening map to two
extragalactic reddening measurements.  These tests provide
confirmation that our dust maps are indeed suitable estimates of 
reddening, as good as or better than those of BH.
These tests also normalize the amplitude of
reddening (in magnitudes) per unit of $100\micron$ flux (in $\MJypSr$).
Our reddening estimates can be written simply as
\BE \Ebv = p~\NDmap = p~\Icorr \Xmap , \EE
where we seek the calibration coefficient $p$.
$\NDmap$ represents the point-source-subtracted, IRAS-resolution
$100\micron$ map, corrected to a reference temperature of $18.2\K$ using
the DIRBE temperature map (equation \ref{equ_NDmap_iras}).

The reddening of external galaxies allows a straightforward calibration
of our maps.  For example, consider
the sample of brightest cluster ellipticals of Postman \& Lauer
(1995), for which $\BminusR$ colors are provided for 106 galaxies. These
objects are at modest redshift ($z < 0.05$) and well-distributed
over the sky.
We use the $k$-corrected
colors provided and apply our dust maps to estimate
the reddening toward each galaxy, assuming $\Ebr = 1.64~\Ebv$ for
normal grains with $\Rv=3.1$ (see Appendix \ref{section_filters}).
For a good normalization of the dust map, one expects no
correlation between intrinsic $\BminusR$ for the galaxies and $\Ebv$.
Because the distribution of the
intrinsic $\BminusR$ of the BCG galaxies is not Gaussian, it is best to use a
non-parametric statistical procedure to test the correlation.  We measured
the Spearman rank correlation coefficient between extinction-corrected
$\BminusR$ and $\Ebv$.
For calibration constants $p < 0.0118$ or $p > 0.0196$, the chances that a
random distribution would have a correlation as large as observed is
less than 5\%.  Thus, our 95\% confidence limits for the normalization of
the maps is $p=0.016 \pm 0.004$.  As confirmation of
the procedure, we applied the same statistic to these galaxies using
the BH reddening estimates:
\BE \Ebv = q~\EbvBH . \EE
$\EbvBH$ represents the BH reddening in $\BminusV$, limiting the analysis
to the 99 galaxies with BH estimates.
For a scaling factor in the range 
$0.66 < q < 1.05$, the BH reddening applied to the BCG galaxies leave no
significant correlation between intrinsic $\BminusR$ and foreground reddening, 
indicating that the BH normalization ($q=1$) is acceptable.

A sharper test of the reddening calibration is possible by using an auxiliary 
correlate to reduce the variance of the intrinsic color distribution.
For example, the $\Mgtwo$ index described by Faber \etal\ (1989)
is expected to correlate closely with the intrinsic $\BminusV$ color
for elliptical galaxies.  For both increased metallicity and increased 
stellar population age, the line-strength increases and the colors become
redder.  Use of the the $\Mgtwo$ index is ideal, since it is available for large
samples and is not affected by reddening.  We use the sample of 472 elliptical
galaxies presented by Faber \etal, which has very broad sky coverage,
including many galaxies in rather dusty directions.  Of this sample,
389 galaxies have photoelectric colors in $\le 30\arcsec$ apertures,
$\Mgtwo$-index measurements, and estimates of BH reddening.
Five of these galaxies are reported by D.\ Burstein to have suspect
photometry and have been removed for our list
(NGC 83, 545, 708, 1603, and 7617).
The $\BminusV$ colors are listed in this
catalog as being $k$-corrected and reddening-corrected in a manner that
deviates slightly from BH reddenings given in the published full-sky BH map.
In one region of the sky ($230\degree < l < 310\degree$,
$-20\degree < b < +15\degree$) it was determined that the BH map is
not reliable and reddenings were assigned based upon deviations in the
Mg-color relation (Burstein \etal\ 1987).  We have added the tabulated
reddening corrections back to the colors to obtain the raw
(only $k$-corrected) colors before proceeding with our analyses.
We should note that the unreliable region lacks a dust-to-gas
ratio estimate in the BH map construction.

Dave Burstein kindly sent us an updated version of Faber \etal's Table 1,
in which the $\Mgtwo$ estimates have been improved. 
For these objects, we compute a linear regression of reddening-corrected
$\BminusV$ against $\Mgtwo$, with residuals $\delta\BminusV$.
We compute the Spearman rank correlation coefficient of $\delta\BminusV$
versus the reddening estimate, again arguing that a good dust map will
have no residual correlation.  For the BH reddenings as listed by
\cite{faber89}, or for the published BH reddening estimates,
there is no significant residual correlation between 
$\delta\BminusV$ and the reddening estimate.  Our formal normalization
of the BH map is $q = 0.90 \pm 0.09$, with the standard deviation
in $\delta\BminusV$ reduced from $0.048\MAG$ to $0.031\MAG$.

\begin{figure}
\plotone{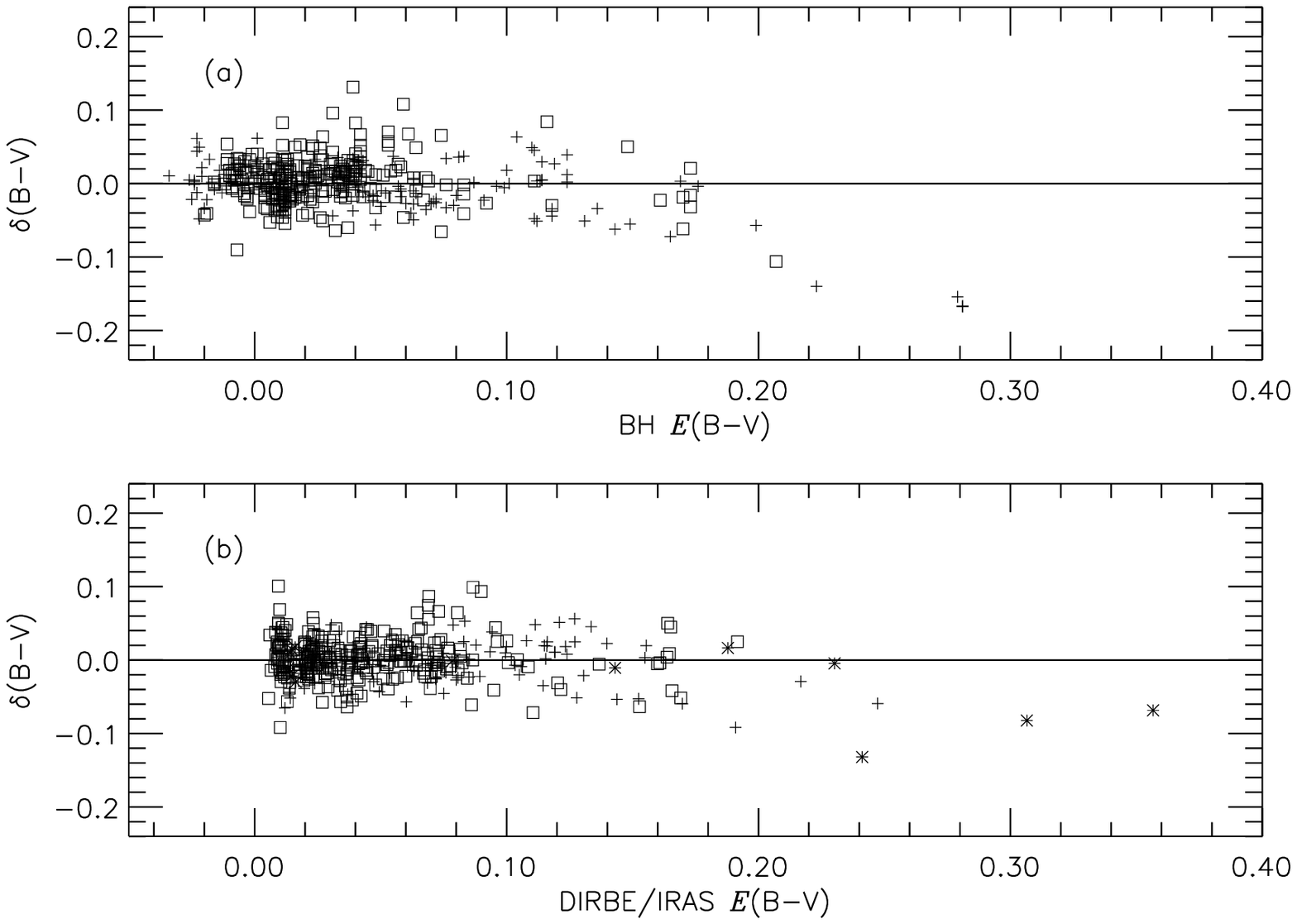}
\caption{Residuals from the $\BminusV$ vs.\ $\Mgtwo$ regression, plotted
  against foreground reddening, for (a) the BH maps and (b) the DIRBE/IRAS maps.
  Plus symbols represent galaxies at southern declinations where the BH maps
  lack dust-to-gas ratio information, and asterisks are those lacking any
  BH values.
}
\label{fig_Mgcolor}
\end{figure}

\placefigure{fig_Mgcolor}

Applying the same procedure to the DIRBE/IRAS maps we find a 95\% confidence
range $p=0.0184 \pm 0.0014$ for the calibration constant.
For $p$ values in this range, the standard deviation of $\delta\BminusV$
is reduced to $0.028\MAG$.
A plot of $\BminusV$ residuals from the Mg-color relation
versus reddening is shown in Figure \ref{fig_Mgcolor}, using both BH
and DIRBE/IRAS reddening estimates.  A different symbol type is used
for galaxies in the north (DEC$>-23\degree$; squares) versus the south
(DEC$<-23\degree$; crosses), as the BH maps do not utilize dust-to-gas
information for the southern points.  Galaxies with no BH map values
were not used in the fits, but are shown in the lower panel as asterisks.
A slight trend in the residuals is evident for both BH and DIRBE/IRAS
corrections, in the sense that the highest reddening values appear to be
overestimated.  However, this trend is not statistically significant
for the DIRBE/IRAS corrections.  Note that the galaxies lacking BH
estimates (asterisks) were not used in the fits, but still have reasonable
color residuals using DIRBE/IRAS reddening corrections.

The larger size and reduced intrinsic-color scatter of the Faber \etal\ 
sample leads to a much tighter constraint on $p$, one which is consistent
with the value derived from the BCG sample.  The $10\%$ precision of the
calibration is remarkably tight, and is consistent with early estimates of
IRAS reddening calibrations from Rowan-Robinson \etal\ (1991).
(However, we note that IRAS-only estimates of the reddening severely
suffer from the poorly-constrained zero-point, zodiacal contamination,
and un-accounted variations in the dust temperature.)

The accuracy of the reddening maps can be estimated from the residuals
in the Mg-color relation.  We assume that the errors in the reddening
estimates, $\sigDI$, are fractional, increasing with increasing reddenings
such that
\BE \sigDI = f~\Ebv . \EE
We further assume that the intrinsic dispersion in the Mg-color relation
is $\sigBV=0.0257\MAG$ (including measurement errors),
which is the measured dispersion for a sub-sample of
galaxies in clean regions of the sky.  A total error for the color of
each galaxy is the quadrature sum of $\sigDI$ and $\sigBV$.
Using this model for the errors, one can determine the accuracy of
the reddenings by increasing $f$ until the $\chi^2$ of the fit equals
the number of degrees of freedom.  For the BH reddening maps, we find
$f=0.29$, and for the DIRBE/IRAS reddening maps we find $f=0.16$.
This demonstrates that the DIRBE/IRAS reddening estimates have an accuracy
of $16\%$, which is nearly a factor of two improvement over the BH estimates
for this full-sky comparison.
In regions of low reddening, $\Ebv \simlt 0.1\MAG$, this data set indicates
that DIRBE/IRAS and BH reddening estimates may be equally good.

The Burstein \& Heiles reddening map appears to be more accurate in
the Northern sky ($\delta_{1950} > -23\degree$), where
where galaxy counts were used to model gas/dust variation.
Splitting the data into two subsamples,
one finds $f=0.25$ for the $257$ galaxies at $\delta_{1950} > -23\degree$,
and $f=0.31$ for the $115$ galaxies at $\delta_{1950} < -23\degree$.
Burstein \etal\ (1987) identify one region in the Southern sky,
approximately bounded by $l=[230\degree,310\degree]$, $b=[-20\degree,20\degree]$,
where a low dust/gas ratio is presumed to make the BH predictions too large.
We find that the dust/gas ratio in this region is variable, but not
unusual on average.  Comparing the BH map to Leiden-Dwingeloo 21-cm data
in the South, we conclude that the BH map may be unreliable near their
survey edges at $b = \pm 10\degree$.  Within the region
identified by Burstein \etal, we find a smaller region,
bounded by $l=[230\degree,240\degree]$, $b=[-15\degree,-10\degree]$,
which appears corrupted and too high in the BH map.

We had originally attempted to use counts-in-cells of the APM galaxy survey
(Maddox \etal\ 1990b, 1990c) as a normalization of the dust map.
Extinction as measured by the
DIRBE/IRAS column-density maps can be calibrated by studying the
statistical covariance between the APM and DIRBE/IRAS dust maps: dusty regions
have increased dust emission and diminished galaxy counts.
However, we encountered a problem similar to that
seen by Heiles (1976) when he attempted to compare reddening measures to the
counts of Shane-Wirtanen galaxies.  Using the APM counts,
we find a normalization ($p$) that is
approximately twice that described above.  The extra sensitivity
of the galaxy counts to dust is almost certainly due to the catalog of
galaxies being surface-brightness limited as well as magnitude limited;
increased foreground dust not only diminishes the total flux of the galaxy
but diminishes the size of the isophotal aperture and increases the likelihood
that the galaxy will be either classified as a star or not be counted at all.
We shall provide details of this analysis in a separate paper
(\cite{finkbeiner97}).

\section{Processing Summary}
\label{section_psummary}

We have extensively processed the available far-IR data sets to generate a 
uniform-quality, column-density map of the dust that radiates from
$100\micron$ to $240\micron$.
The major steps in the processing are summarized below:
\begin{enumerate}
\item The annual average DIRBE maps have been regenerated such that
  the same zodiacal dust column is observed at all wavelengths.
  Data with solar elongation $e < 80\degree$ has been deleted.
\item A zodiacal light model has been constructed from the DIRBE $25\micron$
  map.  The model parameters were constrained by forcing the dust
  to trace the gas (as measured by the Leiden-Dwingeloo \HI\ survey)
  at high Galactic latitudes.
\item Striping artifacts in the IRAS/ISSA $100\micron$ maps
  were removed using a Fourier-space filtering algorithm.
\item Asteroids and non-Gaussian noise was removed from the IRAS/ISSA maps
  using a de-glitching algorithm that compared multiple scans.
\item The IRAS and DIRBE $100\micron$ maps were combined, preserving
  the DIRBE zero-point and calibration.
\item Stars and galaxies were removed to a limiting flux of 
  $f_{60}= 0.6\Jy$ over most of the sky at $|b| > 5\degree$,
  and parts of the sky at lower latitudes.
\item The color temperature of the dust was derived from the DIRBE
  $100\micron$ and $240\micron$ maps.  A temperature-correction map
  is used to convert the $100\micron$ cirrus map to a map proportional
  to dust column.
\item The temperature-corrected dust map is calibrated to the reddening
  of external elliptical galaxies.
  The use of the $\BminusV$ vs.\ $\Mgtwo$ correlation
  allows us to set the calibration constant to 10\% precision.
\end{enumerate}
We have confirmation of the superiority of the new procedure
by a test of the scatter of residual $\BminusV$ colors after regression
against $\Mgtwo$ line-strength.  The scatter of the regression is dominated
by intrinsic scatter among the galaxies.  We measure a reduction in
the variance consistent with a fractional error in the reddening estimates
of $16\%$ for the new maps versus $29\%$ for the BH estimates.


\placefigure{plate_slice}

\section{Discussion}
\label{section_discussion}
 
\subsection{The Nature of the Full-Sky Infrared Cirrus}

The new data show a wealth of filamentary structure that extends to the
smallest scale resolved by our maps.  A representative slice of the
sky is shown in Figure \ref{plate_slice}, where we present the BH
image, the Leiden-Dwingeloo \HI\ image, the dust map with DIRBE resolution,
and the final dust map with IRAS resolution.  This slice displays part
of the Polaris Flare on the left and the delicate structure of the
high-latitude Draco molecular cloud complex (\cite{herbstmeier93};
\cite{mebold85}), also known as MBM complex 26 (\cite{mbm85}), to the right of
center. 
The small-scale structure in the plot is at
the resolution limits of the IRAS data.  Higher resolution data
would perhaps show even finer scale structure.

\placefigure{plate_dustmap}

The full-sky dust map is shown as Figure \ref{plate_dustmap} with a
logarithmic stretch to emphasize the faint, high latitude features.
The cirrus filaments arch across the Galactic poles, and between these
filaments there are several low-density holes.  The Lockman hole, the
region of minimum flux in the \HI\ maps ($l=150.5\degree,b=53\degree$;
\cite{lockman86}), is one such low density region, with a
temperature-corrected flux of $0.39\MJypSr$.  The Southern Galactic
sky contains regions with even lower dust emission, perhaps by a
factor of two.  The lowest column density holes are located at
$l=346.4\degree, b=-58.0\degree$, and $l=239.7\degree, b=-48.6\degree$,
with temperature-corrected fluxes of $0.18\MJypSr$.  Unfortunately, no
high-quality \HI\ data exists to verify that the lowest-density holes
are also minima in \HI\ emission.  A tabulation of dust properties for
our low-density holes and the Galactic poles appears in Table
\ref{table_holes}.  A large region of the Southern Galactic sky
($b<-40\degree, 160\degree < l < 320\degree $) appears to have very
low dust emission and should therefore have the least foreground
contamination for CMBR studies or large-scale-structure analyses of
redshift surveys.  Projects in the northern hemisphere, such as the
Sloan Digital Sky Survey and the North Galactic strip of the Two
Degree Field (2dF) survey, may be more compromised by Galactic extinction.
In particular, the North Galactic strip of the 2dF is significantly
more dusty than their Southern region.
However, the corrections derived from the new dust maps should prove
adequate for most analyses.
 
\placetable{table_holes}

\begin{figure}
\plotone{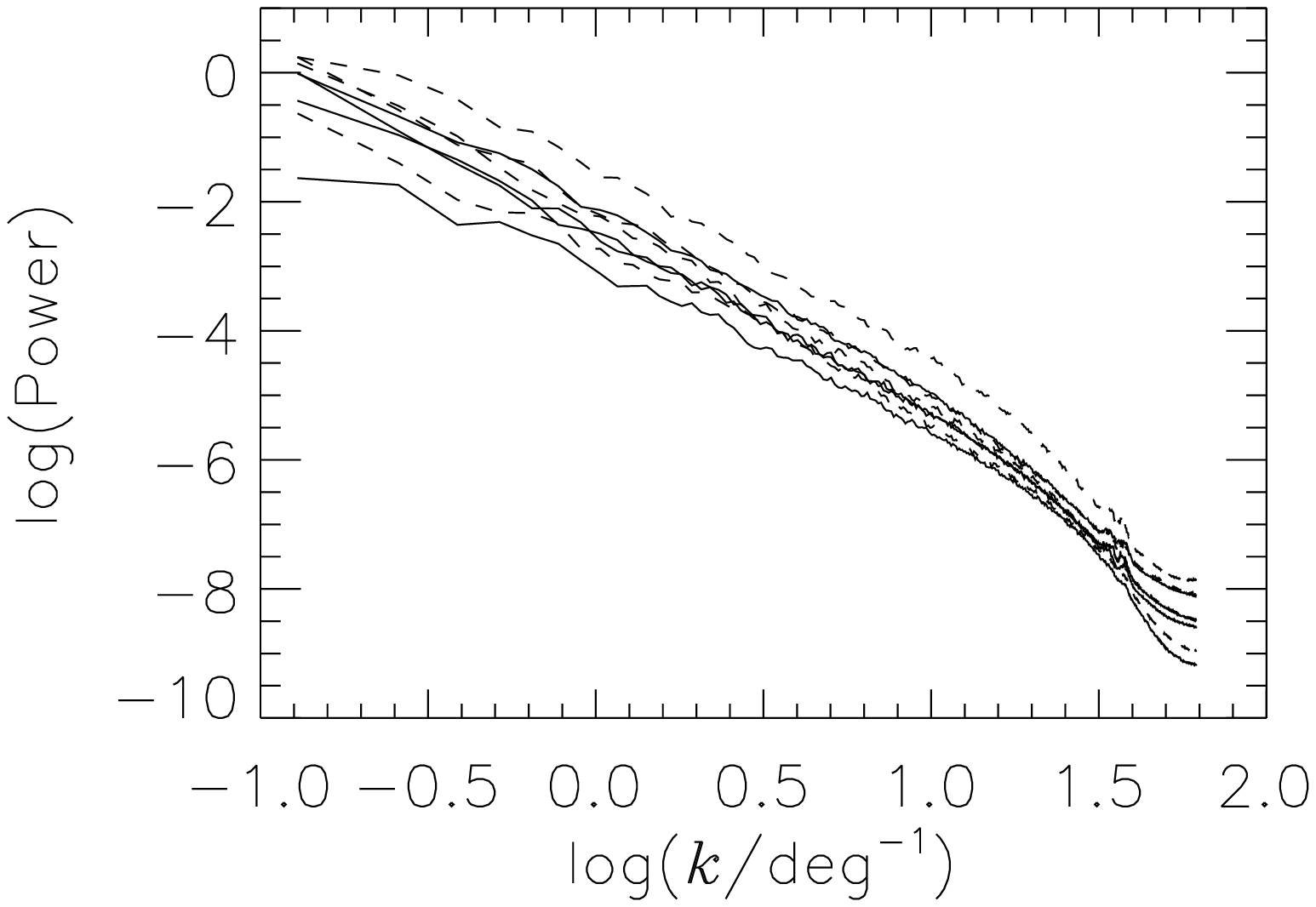}
\caption{Power spectrum of dust.  The four solid curves represent four
  quadrants in the North Galactic sky at $|b| > 45\degree$, while the
  four dashed curves represent four quadrants of the South Galactic sky.}
\label{fig_pkdust}
\end{figure}

\placefigure{fig_pkdust}

In Figure \ref{fig_pkdust}, we show the power spectra of intensity fluctuations
of the dust map calculated separately in eight
independent patches of sky at high Galactic latitude.  For each
hemisphere, we have extracted four quadrants in Galactic latitude bounded
by $|b| > 45\degree$.
The power spectra of these high-latitude patches
have no preferred scale, but are reasonably-well described as power laws
with $P(k) \propto k^{-2.5}$.  The amplitude of the power is
different in each patch, which is indicative of extreme
phase-coherence in the cirrus structure of the Galaxy. To assume that 
the fluctuations can be approximated as random phase power
is a completely inadequate description of this infrared cirrus, and could
lead to misleading estimates of Galactic foregrounds in
CMBR experiments.

\begin{figure}
\plotone{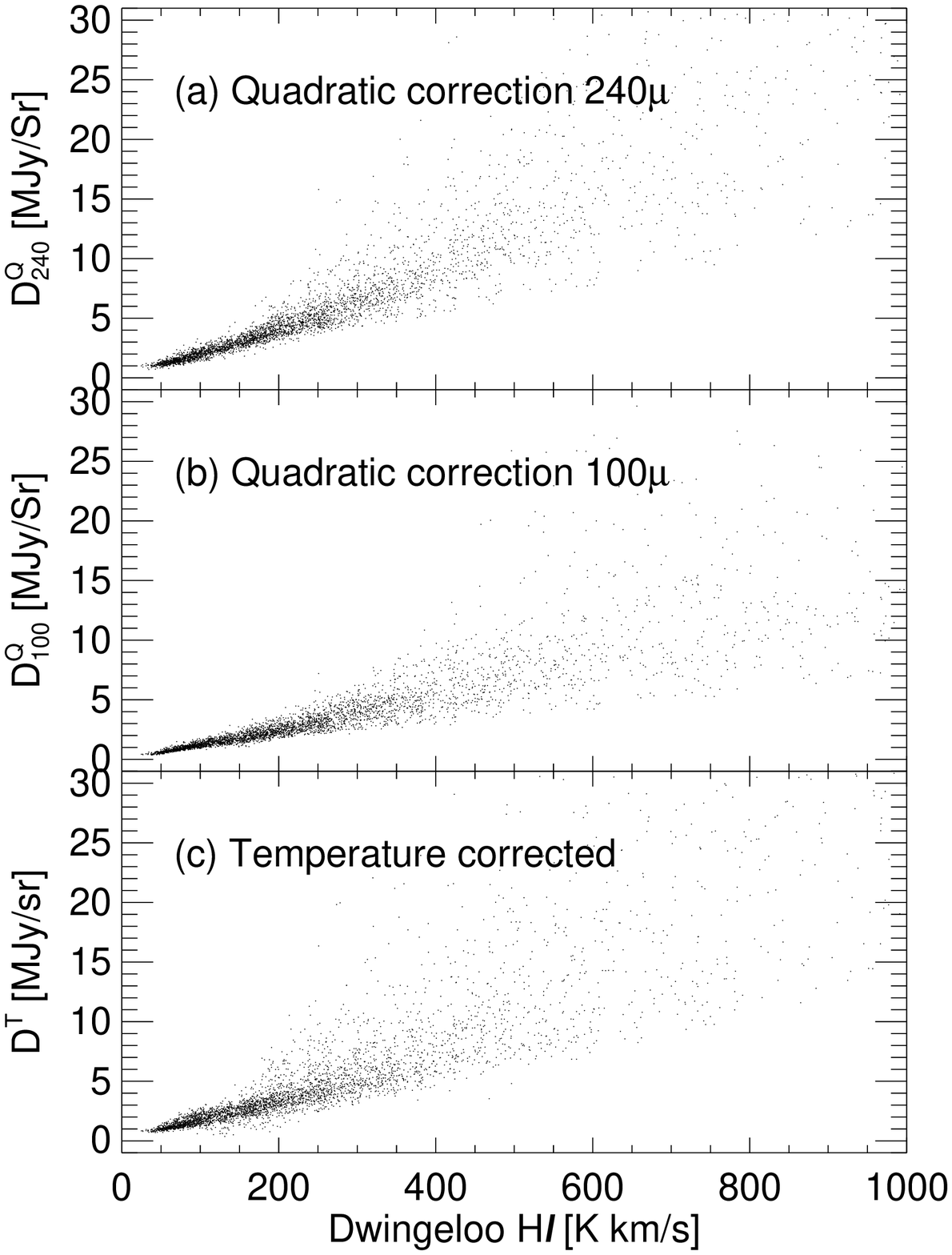}
\caption{The \HI\ correlation with (a) DIRBE $240\micron$ (corrected for
  zodiacal contamination), (b) DIRBE $100\micron$ (also corrected), and
  (c) our derived dust column density.
  This plot demonstrates that the gas to dust relationship deteriorates
  at high flux levels.}
\label{fig_HIdust}
\end{figure}

\placefigure{fig_HIdust}

\subsection{Differences between the Dust and \HI\ maps}

As seen in Figure \ref{fig_HIdust}, the correlation between the dust and the
\HI\ emission is remarkably tight for low flux levels, but has substantial
scatter at higher flux levels.  Because the $240\micron$ map is on the
Rayleigh-Jeans portion of the dust emission curve, one might expect it to
better correlate with the \HI\ than the $100\micron$ map.
We see the opposite effect: there is more
scatter in the $240\micron$ - \HI\ correlation.  We attribute this partly to
the considerably larger noise of the $240\micron$ map.  But some of this
increased scatter is undoubtedly generated by cool molecular cloud regions, in
which much of the hydrogen is in molecular form, and the $100\micron$ is
exponentially reduced because of the lower temperature, while the $240\micron$
emission is suppressed only linearly with temperature.  Furthermore, the
dust-\HI\ scatter is non-negligible even in the regions of low flux, where the
gas is expected to be predominantly neutral.  This scatter may indicate 
either fluctuations in the ionization fraction of the gas, extensive regions
of $\Hplus$, or variations in the dust-to-gas ratio within neutral zones.

\placefigure{plate_dust_gas}

A ratio map of temperature-corrected dust to the Leiden-Dwingeloo \HI\
map is shown as Figure \ref{plate_dust_gas}.  The Leiden-Dwingeloo map
includes \HI\ gas from the full velocity range of the survey ($-450
\leq \vLSR \leq +400\kms$) converted to dust column via the ratio
$0.0122~(\MJypSr)/(\Kkms)$
found in Table \ref{table_zodycoeff}.  Note that there are fluctuations in the
ratio which are spatially coherent, indicating they are not instrumental
noise, but rather real features on the sky.  Most of the high-latitude sky
is a neutral grey, while regions of high dust-to-gas ratio appear as white.
The most prominent of these features at high latitudes must be diffuse
interstellar molecular clouds as well as regions of saturated \HI\
emission, such as Orion ($l \approx 180\degree$,$ b \approx
-20\degree$) and Ophiuchus ($l \approx 0\degree$, $b \approx 15\degree$).
The Polaris Flare ($l \approx 150\degree$, $b \approx 40\degree$) is
very conspicuous.  Comparison of Figure \ref{plate_dust_gas} to Figure
\ref{plate_Tmap} shows that these regions of high dust-to-gas ratio often
coincide with the regions of lower temperature, which is as expected,
since the regions are optically thick in the UV and therefore shielded
from the full ionizing flux of the Galaxy.  Although searches for CO
have failed to detect an abundance of molecular gas above the Galactic
disk (e.g.\ Hartmann, Magnani, \& Thaddeus 1997), it is very likely
that molecular hydrogen is very abundant in these cooler regions
(\cite{spaans97}).

Regions of low
dust-to-gas ratio appear as black.  These regions are primarily distant,
high-velocity \HI\ clouds in which the dust, if present, is not as
well-illuminated by a UV radiation field, and is therefore too cold to
emit at $100\micron$.  The Magellanic Stream is partially present in
the South, while numerous high-velocity clouds are very prominent at
high latitude in the North.  

A general gradient of dust-to-gas ratio from
Galactic center to anti-center is perhaps present in the Northern sky,
but if present in the Southern sky, the gradient is obscured by the
Orion nebula.  Even in the grey, high latitude regions, the
dust-to-gas ratio map is not uniform and exhibits fluctuations of $\pm 15$\%
amplitude that are coherent over scales of 10 degrees.  These
modulations might indicate real gas to dust variation or they may hint
at some unresolved instrumental problems.  For example, three
parallel bands in the North are residuals of the imperfect zodiacal
light removal, but their total modulation is only $15\%$
peak-to-trough.

\placefigure{plate_BHdiff}

Figure \ref{plate_BHdiff} is a full-sky map of the DIRBE/IRAS reddening estimate
minus the BH estimate, for the region $|b|>10\degree$.  Recall that the BH
maps are largely \HI\ maps, with the zero-point adjusted and with smooth
variations in dust-to-gas ratio computed on the basis of galaxy counts.  As
expected, apart from an offset, the BH reddening maps are very close to the
new reddening estimates over most high-latitude regions of the sky, rarely
differing by more than $0.02\MAG$ (aside from a global zero-point difference).
But there are large systematic differences
at low latitude and towards molecular clouds, such as Orion and Ophiuchus.
These modulations largely reflect the direct differences in the dust-to-gas
ratio maps (Figure \ref{plate_dust_gas}). These coherent modifications of the
reddening estimates are important for full-sky analyses of the galaxy
distribution, particularly for studies of large-scale flow fields.

\subsection{The Reddening at the Galactic Poles}

\label{section_rpole}

The absolute zero-point of dust column density in the Galaxy is both
difficult to model from 21-cm or infrared emission, and difficult
to measure directly from stellar colors.  The zero-point of 21-cm maps
are uncertain due to sidelobe contamination of the radio telescopes.
Furthermore, the 21-cm maps completely miss any dust associated with
ionized hydrogen.  Although the DIRBE instrument chopped against a cold load 
to insure a stable zero point, the zero point of the combined DIRBE/IRAS dust
maps are uncertain because of uncertain zodiacal contamination 
and a possible isotropic, extragalactic background.
Direct measure of the reddening is an equally difficult task.
Extragalactic sources can not be used because their intrinsic colors
are not known. Studies of Galactic sources
are limited to calibration of the dust column density
between local and halo stars.
Such comparisons may suffer from selection effects, and do not
sample Galactic dust beyond a few hundred parsecs.

The debate as to the reddening zero-point has traditionally focused
on the values at the Galactic poles.  Although we find the absolute
minima in the dust column are not at the poles (see Table \ref{table_holes}),
most studies are near these two lines-of-sight.  Some authors have
claimed that the Galactic poles are essentially free of reddening,
whereas others claim a reddening
of order $\Ebv \approx 0.02-0.05\MAG$, corresponding to an extinction
$\Ab \approx 0.1-0.2\MAG$.
This debate was reviewed by Burstein \& Heiles in 1982, and has
yet to be resolved.

Several large studies have been conducted to measure the colors
of A and F stars near the Galactic poles.  Color excesses from $\ubvby$
photometry for these stars with well-known, intrinsic colors is used
to measure $\Eby$ reddening.  Most of these studies argue for
$\Ebv \simlt 0.01\MAG$.
The data of Hilditch, Hill, \& Barnes (1983) yield an average $\Ebv=0.011\MAG$
for 34 stars at $b>75\degree$ and more than $200\pc$ above the plane.
Furthermore, the subset of these stars in the quadrant $90 < l < 180\degree$
show $\Ebv=0.000\MAG$.
A similar northern hemisphere study by Perry \& Johnston (1982)
is used to argue for negligible reddening for stars within $200\pc$ distance
towards the NGP.
In a companion study in the southern hemisphere, Perry \& Christodoulou (1996)
find $\Ebv=-0.010 \pm 0.016 \MAG$ for 73 stars within $15\degree$ of the SGP.
A reanalysis of some of these data by Teerikorpi (1990) argues these result
are subject to statistical biases, suggesting that the average reddening
reaches $\Ebv=0.04\MAG$ at $400\pc$ above the Galactic plane.

Our estimated reddening, averaged over $10\degree$ in diameter, is
$\Ebv = 0.015\MAG$ at the NGP and $\Ebv = 0.018\MAG$ at the SGP.
This corresponds to a blue-band extinction of $\Ab=0.065$ ($0.080\MAG$)
at the NGP (SGP), using an $\Rv=3.1$ extinction law.
We believe that a nonzero reddening at the poles is compelling for
several reasons.  First of all, we observe both $100\micron$ emission
and \HI\ emission at the poles, not all of which could be scattered
light or zodiacal emission since it is structured and not entirely
smooth at either pole.  The fact that a linear dust-to-\HI\ regression
fits so well through all the low flux regions suggests that this is
truly material in the ISM.  The gas has low velocity so it is
certainly part of the Milky Way, and so therefore must be the dust.
If the A and F star reddenings are not biased by selection effects,
then this gas and dust must be more than several hundred parsecs
above the Galactic plane.

It is theoretically unreasonable for the dust at high Galactic latitude to have
fully evaporated.
Although the dust at the poles might have been preferentially ablated by the
shocks which formed the Local Bubble, such ablation would generally be expected
to act oppositely to the coagulation that occurs in dense molecular regions.
The grain size distribution should tip further toward smaller grains, thereby
lowering the value of $\Rv$ but not eliminating the
small grains which dominate UV and optical extinction.

For many analyses, a change in the
zero-point of dust reddening is irrelevant.  But for studies such as
photometry of distant
supernovae, such an effect is quite important, and its neglect
will lead to a systematic errors in the inferred $q_0$.  This arises because the
extinction is not grey, and more strongly affects the B- or V-band spectra of
nearby SNe than the R- or I-band spectra of distant SNe.

An offset is readily apparent between the DIRBE/IRAS reddening maps and
those of Burstein \& Heiles.  We measure a median difference of
$0.020\MAG$ in $\Ebv$ between the maps at high Galactic latitudes
($|b| > 45\degree$), in the sense that the BH maps are systematically lower.
For the reasons stated above, we believe the DIRBE/IRAS zero-point to be
more reliable.  If one were to shift the DIRBE/IRAS maps down by $0.020\MAG$,
the lowest column-density holes would have unphysical, negative reddenings of
$\Ebv = -0.015\MAG$.

\subsection{Measurement of the CIB}
\label{section_cibr_discuss}

The search for the signature of an Extragalactic Background Light (EBL) has a
long history (\cite{peebles71}; \cite{bond86}; \cite{bond91}; \cite{zepf96}; 
\cite{guiderdoni97}; \cite{franceschini97}). 
Such a background is inevitable and can be estimated based upon
a star formation history, and the corresponding injection of
energy and metals into the interstellar medium at various epochs.
The predicted EBL can be expressed as
\BE \nu I_\nu = 0.007 \rho_Bc^2 (\Delta Z)(c/4\pi)(1+z_f)^{-1} , \EE
where $\rho_B$ is the present baryon density and $\Delta Z$ is the
overall metallicity produced at redshift $z_f$ (\cite{bond86}).
For $\Omega_B h^2 = 0.025$ and $\Delta Z = 0.02$, this
yields $\nu I_\nu = 37/(1+z_f) \nWpMMSr$.  
A high background flux thus favors substantial metal production at
low redshift. The emergent flux will appear either in the
UV/visible/near-IR window if the star formation is not very dusty, or
will appear in the $100-300\micron$ window if the majority of the light in 
star forming regions is reprocessed by dust.

In the discussion of Section \ref{section_cibr_fit} above, we described
our regression of the dust maps on the \HI\ data
while removing the zodiacal foreground and a constant background term.
Table \ref{table_cibr_coeff} lists the derived background coefficients, $B_b$,
in $\MJypSr$, as a function of the limiting ecliptic latitude cut $|\beta|$.
Note from this table that the
inferred $B_b$ values drop as the fits are extended to lower
values of $|\beta|$.  This results from the inadequacy of the linear
model for zodiacal foreground removal, especially for the $100\micron$ data.
But for fits with $|\beta| > 20\degree$, the $140\micron$ and $240\micron$ 
fits stabilize and are significantly different from zero.  The error
bars are dominated by the large covariance between the measurement of $B_b$
and $A_b$, which becomes very extreme at large $|\beta|$ cuts.  The error bars
listed are the 95\% confidence intervals, not including systematic effects. 
From variations of these results with ecliptic latitude cuts, and from
simulations with mock IPD maps, we estimate that systematic errors are
comparable to the tabulated, formal errors: $0.5$, $0.2$, and $0.2\MJypSr$
at $100$, $140$, and $240\micron$, respectively.

Based on these fits, we conclude that there does exist a well-quantified
uniform background radiation at $140\micron$ and $240\micron$.  
We prefer to use the coefficients derived from $|\beta| > 30^\circ$ as the
best compromise between an inadequate model at low $|\beta|$ and insufficient
modelling leverage at high $|\beta|$.  Our results 
are similar, although slightly higher, than those reported previously 
by Boulanger \etal (1996).  The only significant difference between our
procedure and that of Boulanger \etal\ is that we fit
all terms of the linear regression simultaneously. 
In perhaps more convenient units, a surface brightness of $1.0\MJypSr$
translates to $\nu I_\nu = 30$, $20$, and $12\nWpMMSr$ at $100\micron$,
$140\micron$, and $240\micron$, respectively.  The derived fluxes from
Table \ref{table_cibr_coeff} thus translate to an upper limit of $15\nWpMMSr$
at $100\micron$, and detected flux of $32 \pm 13 \nWpMMSr$ 
at $140\micron$, and $17 \pm 4 \nWpMMSr$ at $240\micron$,
where we have included the estimated, systematic error.

We suspect that this isotropic flux 
is either extragalactic or is some sort of foreground directly
related to the DIRBE telescope.   No Galactic source is likely to be 
isotropic. We detect no hint of a dust layer distributed like $\csc|b|$ that
is uncorrelated with the \HI\ distribution, as might be expected from
a uniform disk of ionized gas. 
Since these results derive from comparison to the \HI\ distribution, they
are sensitive to unknown offsets in the \HI-dust relation.  We have
assumed that the Galactic \HI-to-dust relationship has no zero offset, which
leads us to more reddening than predicted by the BH maps.  If we force
our mean high latitude reddening to match the BH maps, then the inferred
CIB would \emph{increase}.

A possible instrumental explanation for the strong signature at 
$140\micron$ and $240\micron$ is given by Hauser (1996).
Although the entire DIRBE telescope was operated at $2\K$, which should
minimize radiation by the telescope mimicking an isotropic background, in these
two channels there is a measurable 
radiative offset induced by JFETs, operating at 70K, 
used to amplify 
the detector signals.  Uncertainty in the correction for this effect is
estimated to be 5 (2) $\nWpMMSr$ at 140 (240) $\micron$
(\cite{fixsen97}), which is considerably less than the signal we measure.

Presuming this flux is indeed a cosmic infrared background, 
the inferred value is somewhat above that 
expected from the integrated star formation and dust reprocessing history
of high redshift galaxies, even if they are shrouded in dust
(\cite{guiderdoni97}).  The measured
background is four times larger than that expected based on an empirical
summation of the observed luminosity density of galaxies (\cite{malkan97}).
The numbers we measure might be consistent with
a strong burst of massive star formation in elliptical galaxies, if they
are sufficiently dusty (\cite{zepf96}).
Puget \etal\ (1996) report an isotropic background
detection from analysis of COBE/FIRAS data in the 400-1000$\micron$
region, and our background measurements appear to be on the high side
of the extrapolation of their measurements.

The EBL derived from the optical fluxes in the Hubble Deep Field (HDF), with
modest extrapolation of the galaxy counts, is approximately $7.5
\nWpMMSr$ (\cite{madau97}), a factor of $\sim 2$ smaller than that
reported here.  After masking detected sources in the HDF, Vogeley (1997)
has shown that the remaining EBL, unless it is truly uniform,
has a surface brightness that is at most a small fraction of
the integrated light of the discrete sources. 
These results, together with recent observations of
the UV continuum slope at high $z$ (\cite{pettini97}), 
suggest that much early star formation was very dust-enshrouded,
reprocessing most of the UV, optical and near-IR photons to the
far-infrared. 
An integrated far-IR background flux of such a magnitude, if correct,  
is a very promising sign for high angular-resolution studies 
of the far-infrared.   Space missions such as FIRST and
ISO, and ground-based submillimeter observations with SCUBA on the JCMT 
and with the MMA will have a tremendous opportunity to resolve this background,
or to demonstrate that it does not actually exist.  

A more detailed analysis of the CIB is currently underway by the DIRBE team
(\cite{hauser98}; \cite{kelsall98}; \cite{arendt98}; \cite{dwek98}).
Instead of our
simple, empirical model of the zodiacal foreground, they have solved for a 
detailed three-dimensional model of the interplanetary dust.  They provide 
a much more complete analysis of the separation of Galactic emission from the
infrared background, and their results 
will undoubtedly supersede the preliminary measurements reported here.


\section{Summary}
\label{section_conclude}

We have constructed a full-sky map of the Galactic dust
based upon its far-infrared emission.
The IRAS experiment provides high angular resolution at $100\micron$,
whereas the DIRBE experiment provides the absolute calibration necessary
across several passbands to map the dust color temperature and convert
$100\micron$ emission to dust column density.  Point sources and
extragalactic sources have been removed, leaving a map of the infrared cirrus.
This dust map is normalized to $\Ebv$ reddening using the colors
of background galaxies.  The final maps have a resolution of $6.1\arcmin$,
and are shown to predict  reddening with an accuracy of $16\%$.

The new dust map leads to reddening estimates quite consistent
with the Burstein-Heiles maps in most regions of the sky,
with the new maps proving to be twice as reliable.
These new maps are certainly to be preferred in regions of high extinction,
where \HI-based maps suffer from saturation of the 21-cm line or insensitivity
to molecular hydrogen.
Further tests are encouraged to determine the accuracy of our predicted 
reddenings
and extinction.  In particular, the accuracy of the maps
at $|b| < 10\degree$ has yet to be established.

The maps will undoubtedly prove
useful for analyses of current and future CMBR experiments, as well as
a host of Galactic structure studies.  For example, it will be of interest
to determine whether the temperature variations observed at high latitude
are consistent with molecular line observations, and whether better
constraints on the distance to the high latitude molecular clouds can be
obtained.  Molecular line observations of several of the regions of
low dust-to-\HI\ ratio
(i.e., the high velocity clouds) will give information on the metal
abundance in these regions, which might lead to better constraints on
their distance from the plane of the Milky Way.

We show that the dust correlates very well with the available \HI\ maps over
most of the sky.  The ratio of dust-to-gas can be used to flag
molecular clouds on one extreme, and high velocity \HI\ clouds on the
other.  We note that the lowest regions of dust emission occur
in the Southern Galactic sky, in a region where there exist no high-quality
\HI\ maps.  The South Galactic sky has less power in the dust, and should
be preferred for CMBR experiments and large-scale galaxy surveys.

We argue that Burstein \& Heiles (1982)
underestimate reddening by $0.020\MAG$ in $\Ebv$.
Our predicted extinction at the Galactic poles is $\Ab=0.065\MAG$ (North) and
$\Ab=0.080\MAG$ (South) on $10\degree$ scales.  There do exist holes 
that have extinction as low as $\Ab=0.02\MAG$; the most prominent of these
are listed in Table \ref{table_holes}.
We find the lowest-column-density dust holes are in the Southern Galactic sky,
but it is not known if these are also the regions of lowest \HI.
If these selected regions are indeed low in \HI, then they should become
preferred directions for deep imaging and spectroscopy of extragalactic
sources in the soft X-ray bands, where Galactic extinction is a severe problem.

In the process of generating these maps and removing the zodiacal foreground,
we have detected what appears to be a significant far-IR background flux, 
approximately $32$ and $17\nWpMMSr$ in the $140\micron$ and $240\micron$ 
bands respectively.  This flux is surprisingly high, higher than the integrated
light seen in the Hubble Deep Field.  If our measurement is a detection of
the CIB and not some artifact associated with the DIRBE telescope,
this suggests that early star formation
was heavily dust shrouded in most of the Universe.

The dust maps are publicly available, as described in
Appendix \ref{section_files}.


\section{Acknowledgments}

The COBE data sets were developed by the NASA Goddard Space Flight
Center under the guidance of the COBE Science Working Group and were
provided by the NSSDC.  We thank Sherry Wheelock of IPAC for providing
us with ISSA coverage maps.  We thank Steve Maddox for
supplying the APM maps, Will Saunders and the PSCZ team for providing
positions and identifications from the PSCZ survey,
and Tod Lauer for providing the BCG data.
We are especially indebted to Carl Heiles for stimulating discussions and
invaluable guidance during the course of this project.  Further
discussions with Dave Burstein and Chris McKee were extremely useful.
All the analysis described in this paper was performed in IDL, which
increased our efficiency by an enormous factor.  DPF was partially
supported by an NSF Graduate Fellowship.  This work was supported in
part by NASA grant NAG5-1360.


\newpage
\appendix

\section{Remaining Stellar Contamination}
\label{section_contam}

Stars have been removed from our map to a flux level of $f_{100}\approx 0.3\Jy$
(Section \ref{section_stars}).
At $100\micron$, these sources contribute a total of $26,700\Jy$
at $|b| > 5\degree$, amounting to a mean flux density of $0.0026 \MJypSr$.  
With a median star color of $f_{100}/f_{60}=0.4$, our adopted flux cut
at $60\micron$ roughly corresponds to $f_{100} > 0.3\Jy$.  Stars just
below this limit have a peak flux density of $\sim 0.03 \MJypSr$ at
$100\micron$, corresponding to $\Ab \approx 0.003 \MAG$.  Thus, the
contamination from remaining, individual stars is very small in the
final extinction maps.

\begin{figure}
\plotone{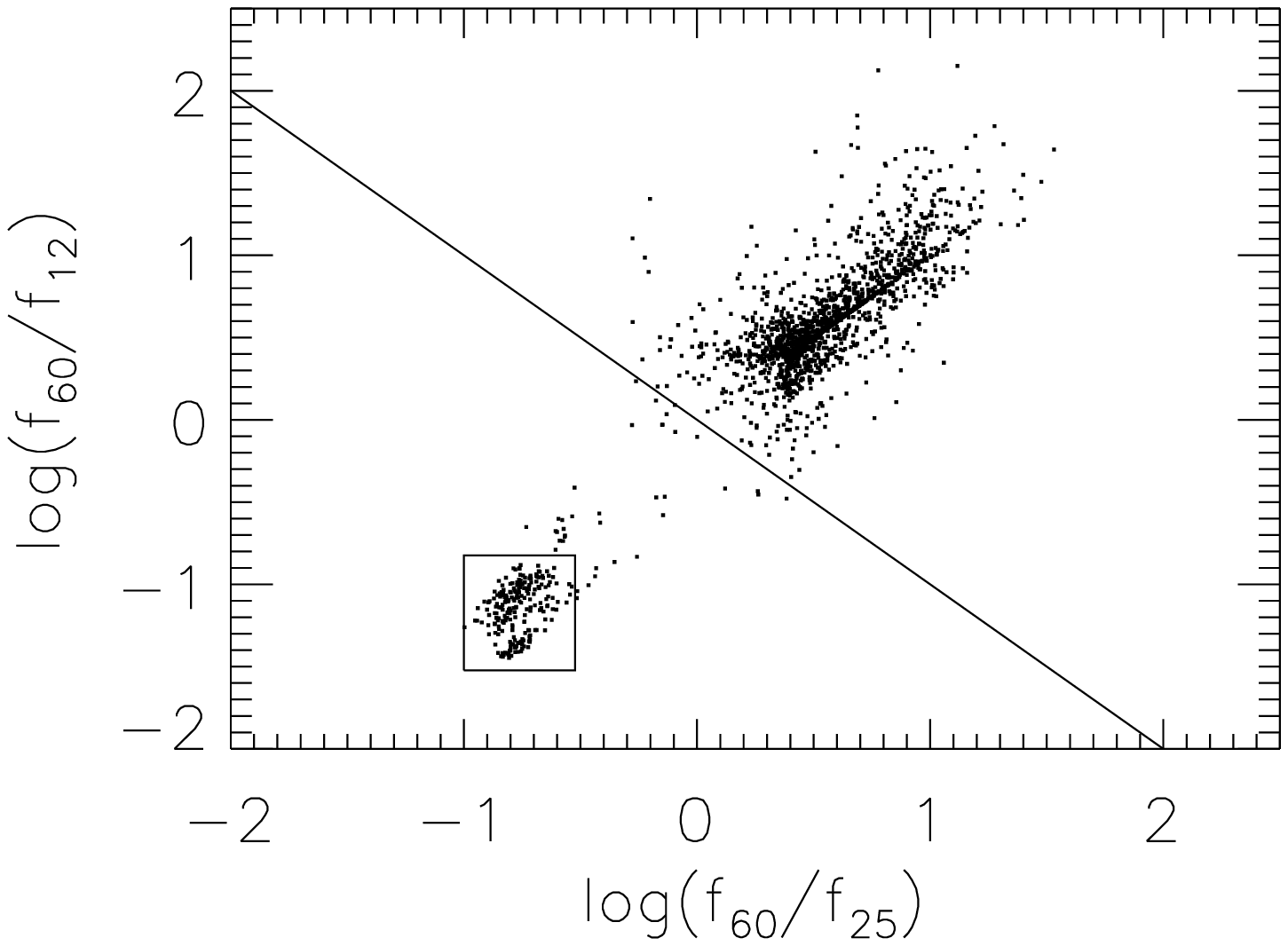}
\caption{Color-color diagram for PSC sources.
  The diagonal line efficiently discriminates between galaxies (above the line)
  and stars (below).  The square box is a strict color cut which retains
  $70\%$ of stars.  For clarity, only 1 in 10 of the stars are plotted.}
\label{fig_stars1}
\end{figure}

\placefigure{fig_stars1}

The concern about remaining contamination from stars is that,
although small on average, it may increase dramatically at low Galactic
latitudes where the star density is very high.
We show that this is not a problem, by extrapolating to the contribution
from stars below our flux limit.  In order to trace the star counts
to low flux and low Galactic latitude, we restrict this analysis to
a strict color cut which limits cirrus confusion:
\begin{enumerate}
\item $ 0.1  < f_{60} / f_{25} < 0.3  $
\item $ 0.03 < f_{60} / f_{12} < 0.15 $
\item $        f_{100}/ f_{60} < 2    $
\end{enumerate}
This strict color cut retains $70\%$ of the stars, while efficiently
excluding the cirrus.  The color-color plane in Figure \ref{fig_stars1}
compares this strict color cut to the looser cut, $f_{60}^2 < f_{12} f_{25}$.
The distribution of stars (satisfying this strict
color cut) as a function of flux is fit by
\BE \Delta\log{N} = 2.37 - 0.84 ( \Delta\log{f_{60}}/0.1 ) . \EE
With a slope in the range $-1<m<0$, the number of stars at faint flux
levels diverges, but the flux from those sources converges.
This allows us to normalize the flux from stars fainter than a given flux
cut, $\fcut$, relative to the number of stars brighter than
that cut:
\BE { {\rm Flux\ of\ stars\ < \fcut} \over {\rm Number\ of\ stars\ > \fcut} }
 = \left( {-m \over 1+m} \right) {\fcut \over \Jy} . \EE
For a flux cut $\fcut=0.6\Jy$, this ratio is $3.2\Jy^{-1}$.
As the number of stars must converge, this ratio represents an
upper limit to the flux from faint stars.
Next, we trace the distribution of stars as a function of Galactic latitude.
Again to minimize cirrus confusion, we limit ourselves to the strict
color cut and brighter stars ($f_{60} > 1.2\Jy$).
The distribution of stars roughly corresponds to a
$\csc|b|$ law for $|b| > 10 \degree$.  At lower latitudes, the PSC is losing
stars primarily due to confusion with cirrus.  At $|b|=10\degree$,
the cirrus has a surface density that is typically 17 times brighter
than a $1.2\Jy$ point source in the ISSA $60\micron$ maps.

\begin{figure}
\plotone{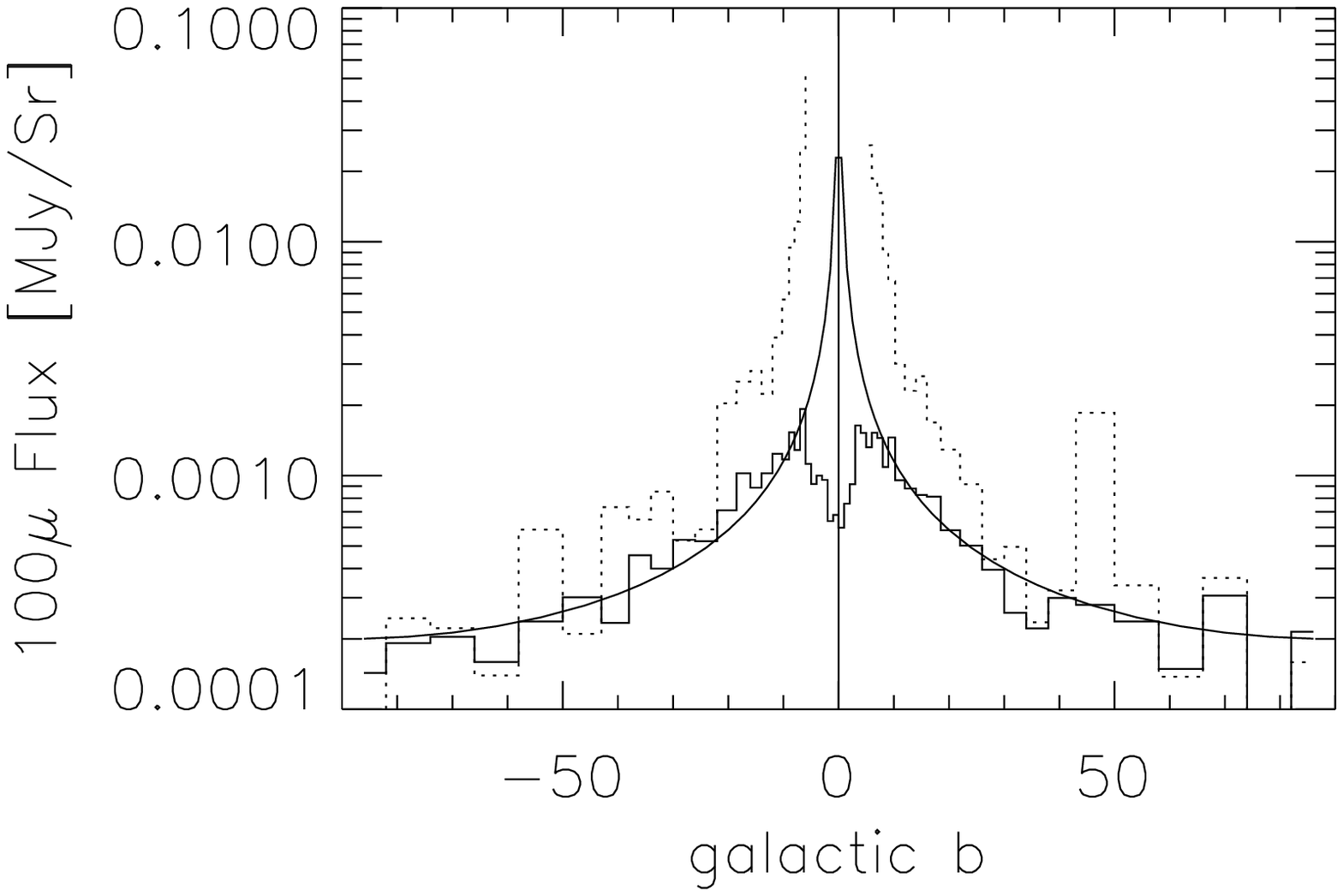}
\caption{Contamination at $100\micron$ from faint stars.
  The solid histogram represents the derived contamination at $100\micron$
  from stars with fluxes below our flux cut.  The dotted histogram shows
  the flux from stars explicitly removed from the maps.
}
\label{fig_stars4}
\end{figure}

We combine the flux and latitude distribution of stars to estimate the
contamination from faint stars.  Using a mean color for stars of
$f_{100}/f_{60} = 0.54$, we plot the estimated contribution from stars
fainter than $f_{60}=0.6\Jy$ in the $100\micron$ maps (Figure \ref{fig_stars4}).
The contamination is approximated by the functional form
\BE F_{100} = 2\times 10^{-4} \csc{|b|} \MJypSr . \EE
For comparison, we overplot the total $100\micron$ flux from the PSC stars
explicitly removed from the maps.
This demonstrates that the faint stars are expected to contaminate the
$100\micron$ maps at a level less than $0.01\MJypSr$ for $|b| > 5 \degree$.
This is not significant, as it represents contamination in derived $\Ab$
values of less than $10^{-3}\MAG$.

\placefigure{fig_stars4}


\section{Extinction in Different Bandpasses}
\label{section_filters}

We have shown that the DIRBE/IRAS dust maps faithfully trace the dust
responsible for reddening of blue light (Section \ref{section_calibrate}).
We need some method of extrapolating the reddening in $\BminusV$
to reddening and extinction in other passbands.
Measuring such relative extinctions is a well-developed industry,
albeit one still fraught with some controversy.
In particular, the composition of interstellar dust is not well known,
nor its variation within the Galaxy.

The selective extinction is variable across optical passbands.
This extinction curve is usually parametrized in terms of the V-band
extinction, $\Av$, and a measure of the relative extinction
between B- and V-band,
\BE \Rv = {\Av \over \Av - \Ab} \equiv {\Av \over \Ebv} . \EE
The value of $\Rv$ varies from $2.6$ to $5.5$ in the measurements
of Clayton \& Cardelli (1988), with a mean value of $3.1$ in the diffuse ISM.
High density regions are presumed to have larger values of $\Rv$ as grains
coagulate into larger grains that are relatively grey absorbers
(\cite{draine85}; \cite{guhathakurta89}; \cite{kim95}).
This variation is certainly a source of concern, as the extinction
curves vary significantly in blue passbands for different values of $\Rv$.
These variations can be well-represented by the same two-parameter function
in any region of the Galaxy.  We use the functional form of O'Donnell (1994)
in the optical and Cardelli, Clayton, \& Mathis (1989) in the ultraviolet
and infrared.  The extinction curve is nearly the same at $\lambda > 8000\Ang$
for all values of $\Rv$, with deviations of up to $30\%$ in V-band and
$55\%$ in B-band.

Fortunately, the observational evidence favors a single extinction curve
for most diffuse clouds in the Galaxy, with $\Rv \approx 3.1$.
Even including stars in dense clouds,
the $76$ measurements of stars from Clayton \& Cardelli show a dispersion
of $0.7$ in $\Rv$.  This translates into a scatter of only $10\%$
in $\Ab/A(\infty)$ and $6\%$ in $\Av/A(\infty)$.  Surprisingly,
Kenyon, Dobrzycka \& Hartmann (1994) find agreement with an $\Rv=3.1$
extinction curve for stars in the heavily-reddened Taurus-Auriga cloud.
Rieke \& Lebofsky (1984) have shown the extinction curve to be
valid at wavelengths $1 < \lambda < 13 \micron$.

The extinction in each passband $b$, $\Delta m_b$, is evaluated
for an extinction in the $\Vband$ passband, $\delta m_V$:
\BEL{equ_extband} \Delta m_b = -2.5 \log{ \left[ {
  \int d\lambda\, W_b(\lambda)\, S(\lambda)\, 10^{-(A_\lambda \Delta m_V/2.5)}
  \over
  \int d\lambda\, W_b(\lambda)\, S(\lambda)
  } \right] } . \EE
The system response, in terms of quantum efficiency, is represented
by $W(\lambda)$, and $S(\lambda)$
is the photon luminosity of the source.
As we have normalized the dust maps to the reddening of elliptical galaxies,
we use an elliptical galaxy for the source.
We average the normal elliptical galaxy SEDs from Kennicutt (1992),
extrapolating the source as $S(\lambda) \propto \lambda$ outside his range
of spectral coverage.
The above expression has been evaluated for a variety of passbands
in the limit of low extinction ($\Delta m_V \rightarrow 0$),
then re-scaled to $\Av = 1$.

The system response is the convolution of the atmosphere, telescope optics,
filter and detector responses.  Where possible, we have chosen the
system responses that correspond to the commonly-used standard stars
for each filter.
A photo-electric Cousins $\UBVRIkc$ system is represented by
the filter responses published by Landolt (1992) convolved
with the quantum efficiency of the RCA 3103A photomultiplier
and the atmospheric transparency at Cerro Tololo.
A CCD Cousins $\UBVRIkc$ system is represented by the filter
responses of the CTIO TEK{\#}2 filter set, convolved with their Tek2K CCD Q.E.
and the atmosphere at Cerro Tololo.
The Stromgren $\ubvby$ system matches the filter curves published
by Crawford \& Barnes (1970), convolved with a Tek2K CCD Q.E. and the
atmosphere at Kitt Peak.
The Gunn $\griz$ system matches the filter curves published
by Schneider, Gunn, \& Hoessel (1983), convolved with the atmosphere
at Palomar.
The Spinrad night-sky filter, $\Rspin$, has been widely used at
Lick Observatory, and is convolved with the Orbit 2K Q.E. and
the atmosphere at Lick.
The infrared $\JHKLp$ bands are represented by the IRCAM3 filter set
at UKIRT, convolved with the atmosphere at Mauna Kea.
The $\ugriz$ filter set for the Sloan Sky Survey represent the total
system response at Apache Point as published by Fukugita \etal\ (1996).
The total system response for some of the broad-band filters for WFPC2
were taken from the Space Telescope Science Institute web page.
The photographic responses for the Second Generation Digitized Sky Survey
are from Weir, Djorgovski, \& Fayyad (1995), convolved with the atmosphere
at Palomar.

The $b_J$ passband used for the APM maps has an unusual definition
and is treated differently.  The APM maps were calibrated
with CCD photometry using the following definition (\cite{maddox90c}):
\BEL{equ_extbj} b_J \equiv B - 0.28 \BminusV . \EE
Combining equations \ref{equ_extband} and \ref{equ_extbj},
one can show the following to be rigorously true:
\BE A(b_J) = 0.72 \Ab + 0.28 \Av . \EE

The results for all passbands appear in Table \ref{table_aratio}.
The effective wavelength, $\lameff$, represents that wavelength on
the extinction curve with the same extinction as the full passband.
The final column normalizes the extinction to photo-electric measurements
of $\Ebv$.  Assuming an $\Rv=3.1$ dust model, the dust maps should be
multiplied by the value in this column to determine the extinction
in a given passband.  Extinctions in narrow passbands can be determined
by evaluating the Cardelli \etal\ or O'Donnell extinction law
at the corresponding wavelength.

\placetable{table_aratio}


\section{Data Presentation}
\label{section_files}

The $100\micron$ map and the dust map are available electronically.
Both maps consist of the destriped, de-glitched ISSA data
recalibrated to DIRBE and point-source subtracted.
The $100\micron$ map maps the emission in units of $\MJypSr$.
The dust map applies our temperature-correction to convert $100\micron$
emission to column density of dust, calibrated to $\Ebv$ reddening
in magnitudes.  Reddenings and extinctions in other passbands are computed
by multiplication with the numbers in Table \ref{table_aratio}.

All maps are stored as FITS files, in pairs of $4096\times 4096$ pixel
Lambert projections.  The NGP projection covers the northern Galactic
hemisphere, centered at $b=+90\degree$, with latitude running clockwise.
The SGP projection covers the southern Galactic hemisphere,
centered at $b=-90\degree$, with latitude running counter-clockwise.
(Note that Figures \ref{plate_Tmap}, \ref{plate_dustmap}, \ref{plate_dust_gas},
and \ref{plate_BHdiff} show the SGP projections rotated by $180\degree$.)
Galactic coordinates ($l,b$) are converted to pixel positions ($x,y$) via
\begin{eqnarray}
  x = 2048 \sqrt{1 - n \sin{(b)}}~\cos{(l)} + 2047.5  \\
  y = - 2048 n \sqrt{1 - n \sin{(b)}}~\sin{(l)} + 2047.5
\end{eqnarray}
where $n=+1$ for the NGP, and $n=-1$ for the SGP.
Pixel numbers are zero-indexed, with the center of the lower left pixel
having position $(x,y)=(0,0)$.
These Lambert projections are minimally distorted at high Galactic latitudes,
with the distortion approaching $40\%$ at $b=0\degree$.
The pixel size of $(2.372\arcmin)^2$ well-samples the FWHM of $6.1\arcmin$.

The caveats to using these maps to measure reddening or extinction
can be summarized as follows:
\begin{itemize}
\item Every effort has been made to remove both extragalactic sources
  and unresolved (Galactic) sources from the dust maps at $|b| > 5\degree$,
  and unconfused regions at lower latitudes.
  Some sources will remain owing to confusion or unusual FIR colors.
  No sources fainter than $0.6\Jy$ at $60\micron$ are removed, although
  these are not significant contaminants.
\item The IRAS satellite did not scan a strip amounting to $3\%$ of the sky.
  In addition, we remove a circle of radius $2\degree$
  centered at $(l,b)=(326.28\degree,+51.66\degree)$ contaminated by Saturn.
  These regions are replaced with DIRBE data, and have no point sources removed.
\item The $100\micron$ passband for the DIRBE satellite is somewhat
  different than the $100\micron$ passband on IRAS.  For a $20\K$
  blackbody, the difference is small, but the color temperature of
  sources is sometimes very different than that of the cirrus.
  This sometimes results in artifacts in the temperature-correction term
  (on scales of one degree).  Our remedy is described in Section
  \ref{section_combine}.
\item The LMC, SMC and M31 are not removed from the maps, nor are sources
  within their Holmberg radius.  Accurate reddenings \emph{through} these
  galaxies is not possible since their temperature structure is not
  sufficiently resolved by DIRBE.  Typical reddenings \emph{towards}
  these galaxies is estimated from the median dust emission in surrounding
  annuli: $\Ebv=0.075\MAG$ for the LMC, $0.037\MAG$ for the SMC,
  and $0.062\MAG$ for M~31.
\item At low Galactic latitudes ($|b| < 5\degree$), most contaminating
  sources have not been removed from the maps, and the temperature structure
  of the Galaxy is not well resolved.  Furthermore, no comparisons between
  our predicted reddenings and observed reddening have been made in these
  regions.  Thus, our predicted reddenings here should not be trusted,
  though inspection of the maps might be of some use.
\item The normalization of the dust column density to reddening has a
  formal uncertainty of $10\%$.
\item Should one wish to change the DIRBE/IRAS dust zero-point to be consistent
  with the Burstein-Heiles maps, we suggest subtracting $0.020\MAG$ in $\Ebv$.
\end{itemize}

The reddening maps will be available in the CD-ROM series of the AAS,
or from the web site {\tt http://astro.berkeley.edu/davis/dust/index.html}.
Mask files are also available that contain the most important processing
steps for any given position on the sky.
Further details will be available with the data files.



\clearpage

\begin{figure}
THIS FIGURE AVAILABLE at http://astro.berkeley.edu/davis/dust/index.html

\caption{Temperature map as determined with an $\alpha=2$ emissivity model.
These Lambert projections are centered on the NGP (top), and SGP (bottom),
with Galactic latitude labelled in degrees.  Lines of constant latitude
and longitude are spaced every 30 degrees.
Note the cool molecular cloud regions and the hot star-forming regions.}
\label{plate_Tmap}
\end{figure}

\begin{figure}
THIS FIGURE AVAILABLE at http://astro.berkeley.edu/davis/dust/index.html
\caption{Fourier destriping of plate 379: (a) raw ISSA HCON-0 image, (b) 
  FFT with wavenumber 0 in center, (c) destriped image, (d)  FFT of
  destriped image.  Note that one of the $k_\theta$ bins contains less
  power than the rest.  These are modes in which power from HCON-3 has
  replaced power in the other HCONs; however HCON-3 covers only half the
  plate, resulting in less power.}
\label{plate_fourier}
\end{figure}

\begin{figure}
THIS FIGURE AVAILABLE at http://astro.berkeley.edu/davis/dust/index.html
\caption{Slice of sky from (a) the BH map, (b) the Leiden-Dwingeloo \HI\ map,
(c) our dust map with DIRBE resolution, and (d) our dust map with
IRAS resolution.  The slice measures approximately $90\degree \times 30\degree$,
centered at $l=100\degree$, $b=+35\degree$.}
\label{plate_slice}
\end{figure}

\begin{figure}
THIS FIGURE AVAILABLE at http://astro.berkeley.edu/davis/dust/index.html
\caption{Full-sky dust map for the NGP (top) and SGP (bottom).}
\label{plate_dustmap}
\end{figure}

\begin{figure}
THIS FIGURE AVAILABLE at http://astro.berkeley.edu/davis/dust/index.html
\caption{Dust-to-\HI\ gas ratio for the NGP (top) and SGP (bottom).}
\label{plate_dust_gas}
\end{figure}

\begin{figure}
THIS FIGURE AVAILABLE at http://astro.berkeley.edu/davis/dust/index.html
\caption{Difference between our reddening map and that of
Burstein \& Heiles (1978, 1982) in the NGP (top) and SGP (bottom).}
\label{plate_BHdiff}
\end{figure}

\clearpage
 
\begin{deluxetable}{rrrrrrl}
\tablewidth{0pt}
\tablecaption{Masked sources in DIRBE maps
   \label{table_dirbebadpix}
}
\tablehead{
   \colhead{$l$}    &
   \colhead{$b$}    &
   \colhead{$\RDmap_{25}$}    &
   \colhead{$\Xmap$}    &
   \colhead{$\Diffmap$}    &
   \colhead{$f_{60}$}    &
   \colhead{IRAS name}    \\
   \colhead{(deg)}  &
   \colhead{(deg)}  &
   \colhead{     }  &
   \colhead{     }  &
   \colhead{     }  &
   \colhead{($\Jy$)}  &
   \colhead{     }
}
\startdata
   4.6 &   35.5 &     & X   &      &       &            \nl
   5.2 &   36.7 &     & X   &      &       &            \nl
  28.8 &    3.5 &     &     & X    &  3041 & 18288-0207 \nl
  49.4 &   -0.3 &     &     & X    &  5263 & 19209+1421 \nl
  80.1 &   -6.4 & X   &     &      &       &            \nl
  90.7 &  -46.5 &     &     & X    &       &            \nl
  94.4 &   -5.3 &     &     & X    &       &            \nl
  95.5 &  -87.5 &     & X   & X    &       &            \nl
 105.0 &   68.8 &     & X   &      &       &            \nl
 106.8 &    5.3 &     &     & X    & 11230 & 22176+6303 \nl
 109.9 &    2.1 &     &     & X    & 12830 & 22540+6146 \nl
 111.5 &    0.8 &     &     & X    &  7073 & 23113+6113 \nl
 118.1 &    5.0 &     &     & X    &       &            \nl
 121.3 &  -21.5 &     & X   &      &       & M31        \nl
 123.1 &   -6.4 &     &     & X    &       &            \nl
 126.6 &   -0.7 &     &     & X    &       &            \nl
 133.5 &  -31.3 &     & X   &      &       &            \nl
 133.7 &    1.2 &     &     & X    & 12000 & 02219+6152 \nl
 141.4 &   40.6 &     & X   & X    &  1170 & 09517+6954 \nl
 165.4 &   -9.0 &     &     & X    &  3121 & 04269+3510 \nl
 168.0 &  -58.0 & X   &     &      &       &            \nl
 173.7 &    2.7 &     &     & X    &  1709 & 05375+3540 \nl
 178.0 &  -31.5 & X   &     &      &       &            \nl
 187.0 &    1.3 & X   &     &      &       &            \nl
 189.9 &    0.5 &     &     & X    &  1157 & 06063+2040 \nl
 197.8 &   56.0 & X   &     &      &       &            \nl
 205.3 &  -14.4 &     &     & X    &  1291 & 05445+0016 \nl
 206.6 &  -16.3 &     &     & X    &  7894 & 05393-0156 \nl
 209.0 &   19.6 & X   &     &      &       &            \nl
 213.7 &  -12.5 &     &     & X    & 13070 & 06056-0621 \nl
 221.4 &   45.1 & X   & X   &      &  5652 & 09452+1330 \nl
 239.4 &   -5.1 & X   &     &      &  1453 & 07209-2540 \nl
 243.2 &    0.3 &     &     & X    &       &            \nl
 265.1 &    1.4 &     &     & X    &  8333 & 08576-4334 \nl
 272.7 &  -39.5 & X   &     &      &       &            \nl
 274.0 &   -1.1 &     &     & X    &  7500 & 09227-5146 \nl
 279.3 &  -33.3 &     & X   &      &       & LMC        \nl
 287.6 &   -0.6 &     &     & X    & 13470 & 10431-5925 \nl
 302.5 &  -44.4 &     & X   &      &       & SMC        \nl
 305.4 &    0.2 &     &     & X    &  8861 & 13092-6218 \nl
 309.3 &   19.2 &     & X   &      &       &            \nl
 316.5 &   21.0 &     & X   &      &       &            \nl
 318.0 &   32.9 & X   &     &      &       &            \nl
 353.1 &   16.9 &     &     & X    &  2196 & 16235-2416 \nl
 359.9 &  -17.7 &     & X   & X    &       &            \nl
\enddata
\tablecomments{List of sources removed from the $\RDmap_{25}$ map used to
construct the zodiacal-light model, the temperature-correction $\Xmap$ map,
or the $\Diffmap$ map used to combine the DIRBE and IRAS $100\micron$ maps.
Sources removed from each map are denoted by an X.  Those that
correspond to a bright source in the IRAS Point Source Catalog have their
$60\micron$ flux and name listed.
}
\end{deluxetable}


\clearpage

\begin{deluxetable}{lrrrrrrr}
\footnotesize
\tablecolumns{8}
\tablewidth{0pt}
\tablecaption{Zodiacal light model coefficients
   \label{table_zodycoeff}
}
\tablehead{
   \colhead{$\lambda$} &
   \colhead{Fit} &
   \colhead{$A$} &
   \colhead{$B$} &
   \colhead{$Q$} &
   \colhead{$\RDmap$/Gas} &
   \colhead{Scatter} &
   \colhead{$Z$} \\
   \colhead{($\micron$)}     &
   \colhead{}    &
   \colhead{}    &
   \colhead{}    &
   \colhead{}    &
   \colhead{($\MJypSr$)/($\Kkms$)} &
   \colhead{(\%)} &
   \colhead{($\MJypSr$)}
}
\startdata
100 & L & $  0.1607\pm 0.0005 $ & $ -0.958\pm 0.015 $ &      ...  &
  $0.01363\pm 0.00012$ &  19  & $ 2.383 \pm .014 $ \nl
140 & L & $  0.0738\pm 0.0005 $ & $  0.058\pm 0.015 $ &      ...  &
  $0.02555\pm 0.00012$ &  21  & $ 1.590 \pm .032 $ \nl
240 & L & $  0.0192\pm 0.0005 $ & $  0.668\pm 0.015 $ &      ...  &
  $0.01895\pm 0.00012$ &  18  & $ 1.068 \pm .023 $ \nl
100 & Q & $  0.0810\pm 0.0020 $ & $  0.623\pm 0.043 $ & 8.79E-4 $\pm$ 2.2E-5 &
  $0.01222\pm 0.00012$ &  16  & $ 2.697 \pm .014 $ \nl
140 & Q & $ -0.0653\pm 0.0020 $ & $  2.82 \pm 0.043 $ & 1.53E-3 $\pm$ 2.2E-5 &
  $0.02309\pm 0.00012$ &  19  & $ 2.132 \pm .032 $ \nl
240 & Q & $ -0.0504\pm 0.0020 $ & $  2.05 \pm 0.043 $ & 7.68E-4 $\pm$ 2.2E-5 &
  $0.01772\pm 0.00012$ &  17  & $ 1.338 \pm .023 $ \nl
\enddata
\tablecomments{$A$, $B$, and $Q$ are the best-fit coefficients for
the linear (L) or quadratic (Q) zodiacal light model.
The ratio between the model-corrected IR-emission and \HI\ emission,
$\RDmap$/Gas, has the quoted scatter in the range $[0,200]\Kkms$.
After removing the contribution from dust,
$Z$ is the zodiacal light plus the CIB averaged at high ecliptic latitudes
($|\beta| > 85 \degree$).
}
\end{deluxetable}
 

\clearpage

\begin{deluxetable}{lrrrrrrrr}
\tablecolumns{4}
\tablewidth{0pt}
\tablecaption{CIB at $100\micron$, $140\micron$, $240\micron$.
\label{table_cibr_coeff}
}
\tablehead{
   \colhead{$|\beta|$ cut}     &
   \colhead{$100\micron$ Flux}      &
   \colhead{$140\micron$ Flux}      &
   \colhead{$240\micron$ Flux}      \\
   \colhead{(deg)}     &
   \colhead{($\MJypSr$)}      &
   \colhead{($\MJypSr$)}      &
   \colhead{($\MJypSr$)}
}
\startdata
   0   & $ -0.96 \pm  0.04$   & $  0.06 \pm  0.08$   & $  0.67 \pm  0.05$  \nl
   5   & $ -0.86 \pm  0.04$   & $  0.28 \pm  0.09$   & $  0.77 \pm  0.06$  \nl
  10   & $ -0.71 \pm  0.04$   & $  0.55 \pm  0.09$   & $  0.92 \pm  0.06$  \nl
  15   & $ -0.49 \pm  0.05$   & $  0.87 \pm  0.10$   & $  1.07 \pm  0.07$  \nl
  20   & $ -0.29 \pm  0.05$   & $  1.13 \pm  0.13$   & $  1.20 \pm  0.08$  \nl
  30   & $  0.02 \pm  0.08$ & {\bf $1.49 \pm  0.20$} & $  1.34 \pm  0.14$  \nl
  45   & $  0.15 \pm  0.19$   & $  1.69 \pm  0.50$   & $  1.27 \pm  0.33$  \nl
\enddata
\tablecomments{Derived CIB level using different cuts in ecliptic
latitude for our zodiacal model.  Errors denote $95\%$ confidence limits,
excluding systematic errors.
The preferred solution uses the cut at $|\beta| = 30 \degree$, as described
in the text.
}
\end{deluxetable}
 

\clearpage
\begin{deluxetable}{lrr}
\tablewidth{0pt}
\tablecaption{Coefficients for $K_b(\alpha,T)$
   \label{table_kfit}
}
\tablehead{
   \colhead{Coeff}           &
   \colhead{$b=100\micron$}  &
   \colhead{$b=240\micron$}
}
\startdata
$a_0$  &    1.00000  &    1.00000   \nl
$a_1$  &    2.18053  & $-$1.55737   \nl
$a_2$  & $-$4.89849  &    0.74782   \nl
$a_3$  &    2.38060  &              \nl
$b_0$  & $-$0.80409  &    0.89257   \nl
$b_1$  &    3.95436  & $-$1.29864   \nl
$b_2$  & $-$4.27972  &    0.61082   \nl
$b_3$  &    1.70919  &              \nl
\enddata
\tablecomments{Coefficients for DIRBE color-correction factors, $K_b(\alpha,T)$,
fit using equation \ref{equ_KofT}.
Fits assume an $\alpha=2$ emissivity model.}
\end{deluxetable}


\clearpage
\begin{deluxetable}{rrrrrrrl}
\tablewidth{0pt}
\tablecaption{Regions of low dust column density
   \label{table_holes}
}
\tablehead{
   \colhead{$\alpha_{2000}$}     &
   \colhead{$\delta_{2000}$}     &
   \colhead{$l$}                 &
   \colhead{$b$}                 &
   \colhead{$\NDmap$}            &
   \colhead{$N($\HI$)$}          &
   \colhead{$\NDmap / N($\HI$)$} &
   \colhead{Comments}            \\
   \colhead{(hr)}                &
   \colhead{(deg)}               &
   \colhead{(deg)}               &
   \colhead{(deg)}               &
   \colhead{($\MJypSr$)}         &
   \colhead{($10^{19} \cm^{-2}$)} &
   \colhead{}                    &
   \colhead{}
}
\startdata
 0 28 & -42 44 & 318.4 & -73.7 &  0.35 &  ... &  ...   &                \\
 0 51 & -27 08 &   0.0 & -90.0 &  0.79 & 15.5 &  0.051 & SG pole        \\
 3 57 & -48 50 & 257.1 & -48.4 &  0.33 &  ... &  ...   &                \\
 3 59 & -42 47 & 248.0 & -49.2 &  0.30 &  ... &  ...   &                \\
 4 01 & -34 25 & 235.2 & -49.1 &  0.20 &  ... &  ...   &                \\
 4 03 & -37 37 & 240.0 & -48.6 &  0.17 &  ... &  ...   &                \\
 4 05 & -35 50 & 237.4 & -48.2 &  0.20 &  ... &  ...   &                \\
 4 44 & -53 20 & 261.3 & -40.2 &  0.23 &  ... &  ...   &                \\
10 36 & +56 38 & 152.7 &  52.0 &  0.29 &  4.4 &  0.066 &                \\
10 48 & +57 02 & 150.5 &  53.0 &  0.44 &  5.8 &  0.077 & Lockman hole   \\
12 51 & +27 08 &   0.0 &  90.0 &  0.66 & 10.1 &  0.065 & NG pole        \\
13 35 & +39 09 &  88.4 &  74.9 &  0.27 &  8.6 &  0.031 &                \\
13 42 & +40 30 &  88.0 &  73.0 &  0.27 &  8.5 &  0.032 &                \\
13 44 & +57 04 & 109.2 &  58.6 &  0.28 & 10.3 &  0.027 &                \\
13 54 & +41 33 &  85.2 &  70.6 &  0.29 &  9.1 &  0.032 &                \\
14 10 & +39 33 &  75.3 &  69.5 &  0.29 &  7.2 &  0.041 &                \\
22 43 & -46 49 & 346.3 & -58.1 &  0.41 &  ... &  ...   &                \\
23 22 & -46 22 & 339.4 & -64.0 &  0.38 &  ... &  ...   &                \\
\enddata
\tablecomments{Coordinates and values for the lowest column-density regions
in the dust map, the Lockman hole, and the Galactic poles.
Multiplication of $\NDmap$ by $0.0184$ recovers the magnitudes of reddening
in $\Ebv$.  \HI\ column densities are for the full velocity range ($-450
\leq \vLSR \leq +400\kms$) of the Leiden-Dwingeloo maps, and are unavailable
for the southern holes.  All numbers are averages over a one-degree aperture.
}
\end{deluxetable}


\clearpage

\begin{deluxetable}{lrrlllrrl}
\footnotesize
\tablewidth{0pt}
\tablecaption{Relative extinction for selected bandpasses
   \label{table_aratio}
}
\tablehead{
   \colhead{Filter}     &
   \colhead{$\lameff$}  &
   \colhead{$\AoverAv$} &
   \colhead{$A/\Ebv$}   &
   \colhead{}           &
   \colhead{Filter}     &
   \colhead{$\lameff$}  &
   \colhead{$\AoverAv$} &
   \colhead{$A/\Ebv$}   \\
   \colhead{}           &
   \colhead{$(\Ang$)}   &
   \colhead{}           &
   \colhead{}           &
   \colhead{}           &
   \colhead{}           &
   \colhead{$(\Ang$)}   &
   \colhead{}           &
   \colhead{}         
}
\startdata
Landolt $U$      &  3372 &  1.664 &  5.434 && Stromgren $u$    &  3502 &  1.602 
&  5.231 \nl
Landolt $B$      &  4404 &  1.321 &  4.315 && Stromgren $b$    &  4676 &  1.240 
&  4.049 \nl
Landolt $V$      &  5428 &  1.015 &  3.315 && Stromgren $v$    &  4127 &  1.394 
&  4.552 \nl
Landolt $R$      &  6509 &  0.819 &  2.673 && Stromgren $\beta$&  4861 &  1.182 
&  3.858 \nl
Landolt $I$      &  8090 &  0.594 &  1.940 && Stromgren $y$    &  5479 &  1.004 
&  3.277 \nl
CTIO $U$         &  3683 &  1.521 &  4.968 && Sloan $u'$       &  3546 &  1.579 
&  5.155 \nl
CTIO $B$         &  4393 &  1.324 &  4.325 && Sloan $g'$       &  4925 &  1.161 
&  3.793 \nl
CTIO $V$         &  5519 &  0.992 &  3.240 && Sloan $r'$       &  6335 &  0.843 
&  2.751 \nl
CTIO $R$         &  6602 &  0.807 &  2.634 && Sloan $i'$       &  7799 &  0.639 
&  2.086 \nl
CTIO $I$         &  8046 &  0.601 &  1.962 && Sloan $z'$       &  9294 &  0.453 
&  1.479 \nl
UKIRT $J$        & 12660 &  0.276 &  0.902 && WFPC2 F300W      &  3047 &  1.791 
&  5.849 \nl
UKIRT $H$        & 16732 &  0.176 &  0.576 && WFPC2 F450W      &  4711 &  1.229 
&  4.015 \nl
UKIRT $K$        & 22152 &  0.112 &  0.367 && WFPC2 F555W      &  5498 &  0.996 
&  3.252 \nl
UKIRT $L'$       & 38079 &  0.047 &  0.153 && WFPC2 F606W      &  6042 &  0.885 
&  2.889 \nl
Gunn $g$         &  5244 &  1.065 &  3.476 && WFPC2 F702W      &  7068 &  0.746 
&  2.435 \nl
Gunn $r$         &  6707 &  0.793 &  2.590 && WFPC2 F814W      &  8066 &  0.597 
&  1.948 \nl
Gunn $i$         &  7985 &  0.610 &  1.991 && DSS-II $g$       &  4814 &  1.197 
&  3.907 \nl
Gunn $z$         &  9055 &  0.472 &  1.540 && DSS-II $r$       &  6571 &  0.811 
&  2.649 \nl
Spinrad $\Rspin$ &  6993 &  0.755 &  2.467 && DSS-II $i$       &  8183 &  0.580 
&  1.893 \nl
APM $b_J$        &  4690 &  1.236 &  4.035 &&                  &       &        
&        \nl
\enddata
\tablecomments{The above tabulates the magnitudes of extinction
evaluated in different passbands using the $\Rv=3.1$ extinction
laws of Cardelli \etal\ (1989) and O'Donnell (1994).
The final column normalizes the extinction to photo-electric measurements
of $\Ebv$. }
\end{deluxetable}


\end{document}